\documentclass[10pt,twocolumn,letterpaper]{article}

\usepackage{cvpr}              % To produce the CAMERA-READY version
\usepackage[dvipsnames]{xcolor}

\newcommand{\defeq}{\coloneqq}

\newcommand{\E}{\mathbb{E}}

\newcommand{\Eb}[2]{\E_{#1}\!\left[#2\right]}

\newcommand{\bx}{\mathbf{x}}

\newcommand{\bepsilon}{{\boldsymbol{\epsilon}}}

\newtheorem{theorem}{Theorem}

\makeatletter
\DeclareRobustCommand\onedot{\futurelet\@let@token\@onedot}
\def\@onedot{\ifx\@let@token.\else.\null\fi\xspace}

\makeatother

\definecolor{yellow}{rgb}{1, 1, 0.7}
\definecolor{orange}{rgb}{1, 0.85, 0.7}
\definecolor{tablered}{rgb}{1, 0.7, 0.7}
\definecolor{red}{rgb}{1, 0, 0}

\definecolor{wincolor}{rgb}{0.85, 0.0, 0.0}

\definecolor{darkyellow}{rgb}{0.8, 0.8, 0.5}
\definecolor{darkred}{rgb}{0.7, 0.3, 0.3}
\definecolor{darkgreen}{rgb}{0.3, 0.7, 0.3}
\definecolor{blue}{rgb}{0.251, 0.498, 0.824}
\definecolor{green}{rgb}{0, 1.0, 0}
\definecolor{pink}{rgb}{1, 0.4, 0.7}

\definecolor{visible-blue}{rgb}{0.286, 0.525, 0.910}
\definecolor{tabfirst}{rgb}{1, 0.7, 0.7} % red
\definecolor{tabsecond}{rgb}{1, 0.85, 0.7} % orange
\definecolor{tabthird}{rgb}{1, 1, 0.7} % yellow

\definecolor{realred}{rgb}{0.95, 0.1, 0.0}

\newcommand{\boldparagraph}[1]{\vspace{0.1cm}\noindent{\bf #1:}}

\definecolor{cvprblue}{rgb}{0.21,0.49,0.74}
\usepackage[pagebackref,breaklinks,colorlinks,allcolors=cvprblue]{hyperref}

\usepackage{pifont}
\usepackage{color}
\usepackage{xcolor}
\usepackage{nicefrac}       %
\usepackage{multirow}
\usepackage{multicol}
\usepackage{dsfont}
\usepackage{mathtools}
\usepackage{amsmath, bm}
\usepackage{colortbl}
\usepackage[percent]{overpic}

\usepackage{tabularx}

\newtheorem{condition}{Condition}

\def\paperID{17451} % *** Enter the Paper ID here
\def\confName{CVPR}
\def\confYear{2025}

\title{Beyond Existance: Fulfill 3D Reconstructed Scenes with Pseudo Details}

\author{Yifei Gao*\\
{\tt\small yilei.jin123@gmail.com}
\and
Jun Huang*\\
\and
Lei Wang$\dag$\\
\and
Ruiting Dai\\
\and
Jun Cheng\\
}

\begin{document}
\maketitle

\begin{figure*}[t]
	\centering
	\includegraphics[width=0.9\textwidth]{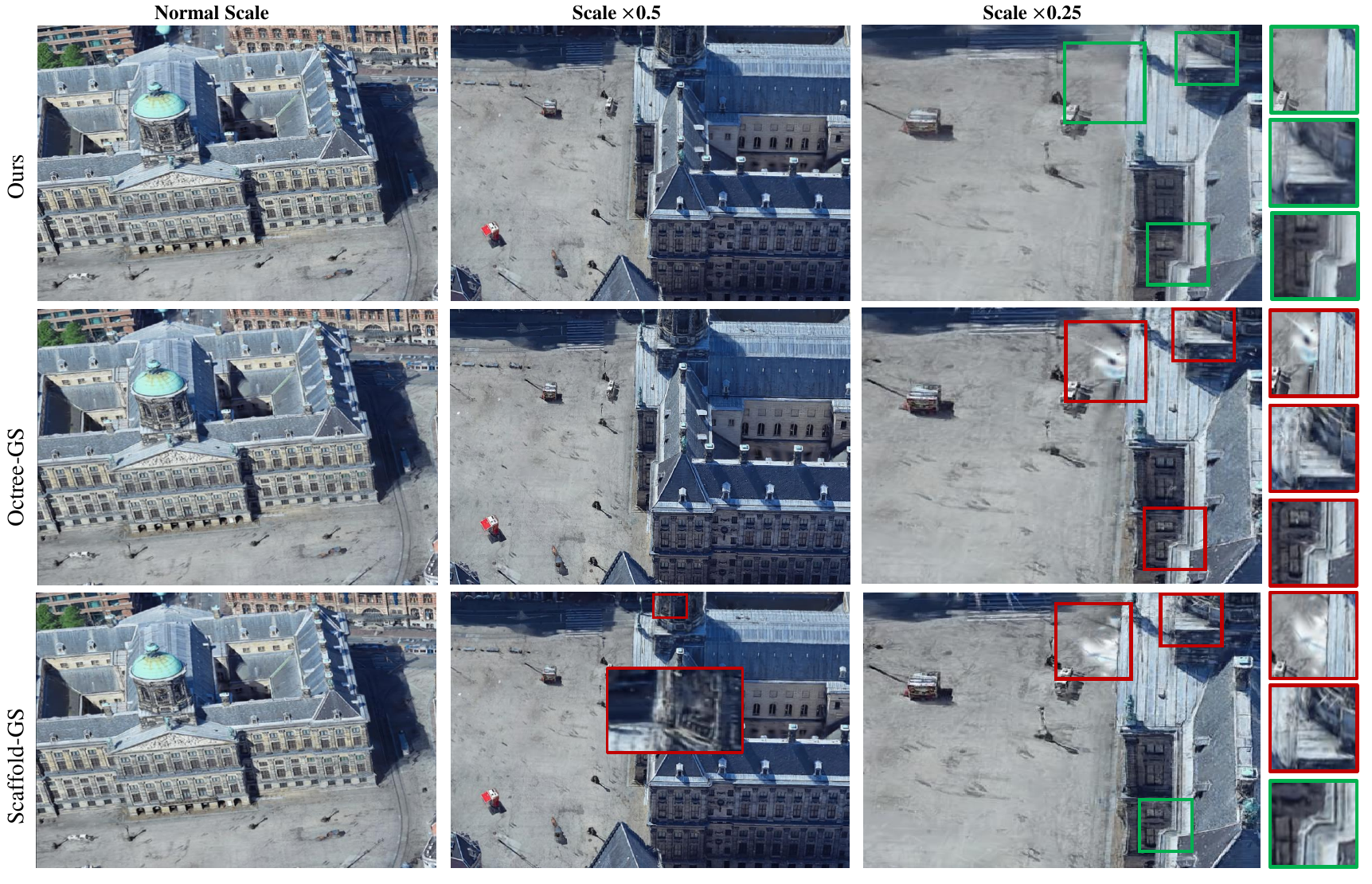}
	%\vspace{-10pt}
	\caption{\textbf{Rendering comparisons.} While most methods render images faithfully at standard scales, their zoomed-in views exhibit significant artifacts that are not apparent at normal magnifications.}
	\vspace{-10pt}
	\label{fig:zoom_artifacts}
\end{figure*}

\begin{abstract}
The emergence of 3D Gaussian Splatting (3D-GS) has significantly advanced 3D reconstruction by providing high fidelity and fast training speeds across various scenarios. While recent efforts have mainly focused on improving model structures to compress data volume or reduce artifacts during zoom-in and zoom-out operations, they often overlook an underlying issue: training sampling deficiency. In zoomed-in views, Gaussian primitives can appear unregulated and distorted due to their dilation limitations and the insufficient availability of scale-specific training samples. Consequently, incorporating pseudo-details that ensure the completeness and alignment of the scene becomes essential. In this paper, we introduce a new training method that integrates diffusion models and multi-scale training using pseudo-ground-truth data. This approach not only notably mitigates the dilation and zoomed-in artifacts but also enriches reconstructed scenes with precise details out of existing scenarios. Our method achieves state-of-the-art performance across various benchmarks and extends the capabilities of 3D reconstruction beyond training datasets.
\end{abstract}

\section{Introduction}
The advent of 3D Gaussian Splatting (3D-GS)~\cite{kerbl20233d} has inspired significant advancements in 3D reconstruction, enhancing a variety of applications such as VR interactions~\cite{jiang2024vr, xie2023physgaussian}, drivable human avatars~\cite{zielonka2023drivable, saito2023relightable, zheng2023gps, qian2023gaussianavatars}, and navigation through large-scale urban scenes~\cite{yan2024street, zhou2023drivinggaussian, lu2023scaffold}. Furthermore, the implementation of 3D-GS in 4D reconstruction~\cite{wu20244dgs, duisterhof2024deformgs, guo2024motionaware3dgs}, splashing effects~\cite{feng2024gaussiansplashing}, style transformation~\cite{liu2024stylegs}, and object segmentation and editing~\cite{ye2024gsgrouping} have emerged, revealing significant commercial potential.

Despite the promising progress of 3D Gaussian Splatting (3D-GS), its relatively straightforward optimization strategies have led to persistent issues such as artifacts in novel-view renderings, redundancy, and limited rendering speed. To tackle these problems, Mip-Splatting~\cite{yu2023mipsplat} utilizes a 3D smoothing filter and a 2D Mip filter to mitigate zoom-in artifacts and zoom-out dilation observed in 3D-GS. Similarly, Mipmap-GS~\cite{li2024mipmapgs} integrates multi-scale pseudo-ground truth training to enhance performance. On another front, Scaffold-GS~\cite{lu2023scaffold} introduces spatial offsets to improve the structure of Gaussians and incorporates a Multi-Layer Perceptron (MLP) for implicit color prediction and generalization. Efforts in compression and quantization~\cite{navaneet2024compact3d, fan2024lightgs, hamdi2024ges} aim to reduce the redundancy of Gaussians while largely maintaining their rendering potential. Octree-GS~\cite{ren2024octreegs} employs an octree structure to enhance Level-of-Detail (LOD) reconstruction, achieving both high rendering quality and volumetric compactness. In contrast to methods that focus on improving modeling techniques, Bootstrap-GS~\cite{gao2024bootstrap3dgs} concentrates on enhancing the sampling process of training data. It enhances training results by bootstrapping novel-view renderings from already trained 3D-GS and reintegrating them into the training sets while maintaining multi-view consistency. 
% emerges as an inspiring approach that 

Despite the progress and contributions of Bootstrap-GS, its training strategies tend to be coarse and time-consuming, and its training process remains precarious when striving to fully alleviate artifacts. Furthermore, while contemporary methods can largely recover scenes in their original views, they struggle to handle zoomed-in renderings using 3D-GS, with most exhibiting significant artifacts in zoomed-in views, as shown in Figure \ref{fig:zoom_artifacts}. Although the filtering approach in Mip-Splatting helps alleviate these issues, it often results in blurred outputs that lack high-frequency details.
% significant

Building upon prior research~\cite{gao2024bootstrap3dgs, ren2024octreegs, yu2023mipsplat}, we introduce a novel training strategy that enhances fine details using upscale diffusion models while ensuring consistency between high-frequency zoomed-in details and low-frequency training data. Additionally, we have developed a refined bootstrapping pipeline to regulate time consumption and minimize fluctuations during training. To validate the effectiveness of our methods, we apply them across multiple frameworks and evaluate their performance on diverse benchmarks.

\noindent In summary, our contributions are as follows:

\begin{itemize}
\item We theoretically demonstrate the feasibility of generating details from a zoomed-in perspective using upscale diffusion models, and achieve excellent results in practical applications. 
\item By integrating upscaling diffusion models during training, our method generates high-frequency, out-of-scene details that remain consistent with the training datasets in standard views, while simultaneously mitigating artifacts in zoomed-in views.
\item We fully exploit bootstrapping techniques and refine its pipeline to achieve state-of-the-art performance across various tasks based on different frameworks, while ensuring a stable and efficient training process.
\end{itemize}

\section{Related Work}
\label{sec:related_work}

\boldparagraph{Image Distortion of Upscaling}
\label{sec:img_distortion}
Scaling adjustments during the training of 3D-GS are achieved by modifying the training camera scales and correspondingly adjusting the resolutions of their ground truth images. For amplified scales, the image resolution is increased using interpolation~\cite{yu2023mipsplat,li2024mipmapgs,chen2022tensorf}. However, when upscaling an image, increasing the sampling rate during interpolation by considering more surrounding pixels cannot overcome the fundamental limit on the highest frequency that can be sampled.

%Traditional image upscaling methods are generally interpolation-based, utilizing colors from surrounding pixels to perform weighted average sampling. Examples include bilinear and bicubic interpolation, which primarily differ in the number of surrounding pixels they consider. 

In practice, to satisfy the constraint of Nyquist-Shannon Sampling Theorem~\cite{nyquist1928certain, shannon1949communication} when reconstructing a signal from discrete samples, a low-pass or anti-aliasing filter is applied before sampling. This filter eliminates any frequency components above half of the sampling rate and attenuates high-frequency content that could lead to aliasing. Consequently, after filtering, the highest signal frequency obtainable from a given image is less than half of the highest original frequency. 

Traditional upscaling methods cannot restore high-frequency details from  filtered signals. In practical applications of upscaling, inadequate sampling can lead to image artifacts such as dilation and aliasing. These issues become particularly significant in regions rich in detail when the upscaling ratio exceeds 2, resulting in noticeable distortions.
%already

%\boldparagraph{Novel View Synthesis} Novel View Synthesis (NVS) involves generating new images from viewpoints different from those of the original captures. Introduced by \cite{mildenhall2021nerf}, Neural Radiance Fields (NeRF) have revolutionized NVS tasks with their photo-realistic rendering quality and view-dependent modeling capabilities with MLP encoding, which maps directly from positionally encoded spatial coordinates and viewing directions. However, NeRF's computational demands are substantial due to intensive MLP forward computations, prompting a significant amount of research focused on optimization and acceleration~\cite{liu2020neural, yu2021plenoctrees, fridovich2022plenoxels, sun2022direct, chen2022tensorf, muller2022instant, xu2023grid, xiangli2023assetfield}. To improve efficiency, Instant-NGP~\cite{muller2022instant} introduces a multi-resolution hash encoding and interpolation-based reconstruction that can be trained within seconds while maintaining high fidelity. TensorRF~\cite{chen2022tensorf} uses 4D tensors to model scenes, decomposing them to enhance storage and rendering efficiency.
\boldparagraph{Mulfi-view Generation by Diffusion Model}
Diffusion models~\citep{sohl2015deep, song2020score, ho2020denoising, nichol2021improved, song2020denoising} are latent variable models using Markov chain to recover the original data distribution $\bx_0$ from an initial random noise. Training is performed by optimizing the standard variational bound on the negative log-likelihood, and the training objective simplifies to~\cite{ho2020denoising}: 
%\begin{footnotesize}
\begin{align}
 L_\mathrm{simple}(\theta) \defeq \Eb{t, \bx_0, \bepsilon}{ \left\| \bepsilon - \bepsilon_\theta(\sqrt{\bar\alpha_t} \bx_0 + \sqrt{1-\bar\alpha_t}\bepsilon, t) \right\|^2} \label{eq:training_objective_simple}
\end{align}
%\end{footnotesize}
where $\epsilon_\theta$ is a learned noise prediction function, $\bar\alpha_t$ is the schedule factor, $t$ is the current time step, and $\epsilon$ is a random constant. 

Diffusion models have been frequently employed as priors in training optimization for 3D reconstruction tasks, particularly in object reconstruction~\cite{chen2024text,liu2023syncdreamer}.  Score Jacobian Chaining is proposed to lift pretrained 2D diffusion model for 3D object generation \cite{wang2023scorejc}. In DreamFusion~\cite{poole2022dreamfusion}, Score Distillation Sampling was introduced to eliminate the need for explicit sampling from diffusion models during training, but only in the latent space. However, DreamGaussian~\cite{tang2023dreamgaussian} demonstrated that this method is ineffective for generating fine-grained details, where exact sampling and further fine-tuning of diffusion models become necessary. Prior to the emergence of Bootstrap-GS, diffusion models were seldom used as priors in large scene reconstruction, primarily because the multi-view alignment of the generated views by diffusion poses significant implementation difficulties.

\newcommand{\RRR}{\mathbb{R}}

\section{Preliminaries}
\subsection{Bootstrap-GS}
\label{sec:boot_intro}
Due to the deficiency in scenario sampling, a bootstrapping technique is employed in~\cite{gao2024bootstrap3dgs} for scenario integration based on 3D-GS. For a brief introduction to 3D-GS, please refer to our \textbf{Supplementary Material}. A novel-view image $\mathbf{I}$ synthesized by the trained 3D-GS model~\cite{kerbl20233d} can be considered a ``partially degraded" version of the ground truth image $\mathbf{I}'$. To approximate the ground truth, a diffusion model (denoiser) $\boldsymbol{\epsilon}_\theta$ with learnable parameters $\theta$ is utilized to perform image-to-image regeneration on the synthesized image, generating a differential $\delta_i$ such that: $\mathbf{I}' \approx \delta_i  + \mathbf{I}.$
\iffalse
\begin{align}
 \mathbf{I}' = \delta_i  + \mathbf{I}.
\end{align}
\fi
The differential $\delta_i$ can be represented as a sum over reverse diffusion processes based on Eq.~\ref{eq:training_objective_simple}:

\begin{equation}
\label{eq:image_intergation}
\delta_i = \sum_{t \in T} \boldsymbol{\epsilon}_\theta\left( \sqrt{\bar{\alpha}_t} \, \mathbf{x}_t(\mathbf{I}, \sqrt{1 - \bar{\alpha}_t} \, \boldsymbol{\epsilon}) \right),
\end{equation}
where $\bx_t$ is a function of $\mathbf{I}$ and $\bepsilon$, $T$ is a set of pre-specified time schedule~\cite{ho2020denoising,song2020denoising}.

To enhance multi-view consistency in regenerated images, Bootstrap-GS employs multiple renderings focused on the same region to perform average sampling. Suitable thresholds are set to ensure faithful cloning and splatting of Gaussians. Given a set of novel-view renderings $\mathbf{I}_n$ and their regenerated counterparts $\mathbf{I}_r$, the bootstrapping loss $\mathcal{L}_{b}$ is defined as the average $\mathcal{L}_1$ loss over each pair of images. Specifically, $\mathcal{L}_{b} = \frac {\lambda_{\text{boot}}}{N} \sum_{i \in N} {\mathcal{L}^i_b}$, where $\mathcal{L}_b^i = \lVert \mathbf{I}_n^i - \mathbf{I}_r^i \lVert$, $\lambda_{\text{boot}}$ is a scalar weighting factor, and $N$ is the total number of variants. By combining this average loss with an appropriate $\lambda_{\text{boot}}$, the bootstrapping loss $\mathcal{L}_{b}$  increases the average sampling steps for the regenerated parts, leading to more faithful re-rendering. For more details, please refer to~\cite{gao2024bootstrap3dgs}.

\section{Methods}

In this section, we first elucidate the theoretical basis of using upscale diffusion models and then illustrate our practical implementations.

%------------------------------------------
\begin{figure}[t!]
\centering
\includegraphics[width=0.8\linewidth]{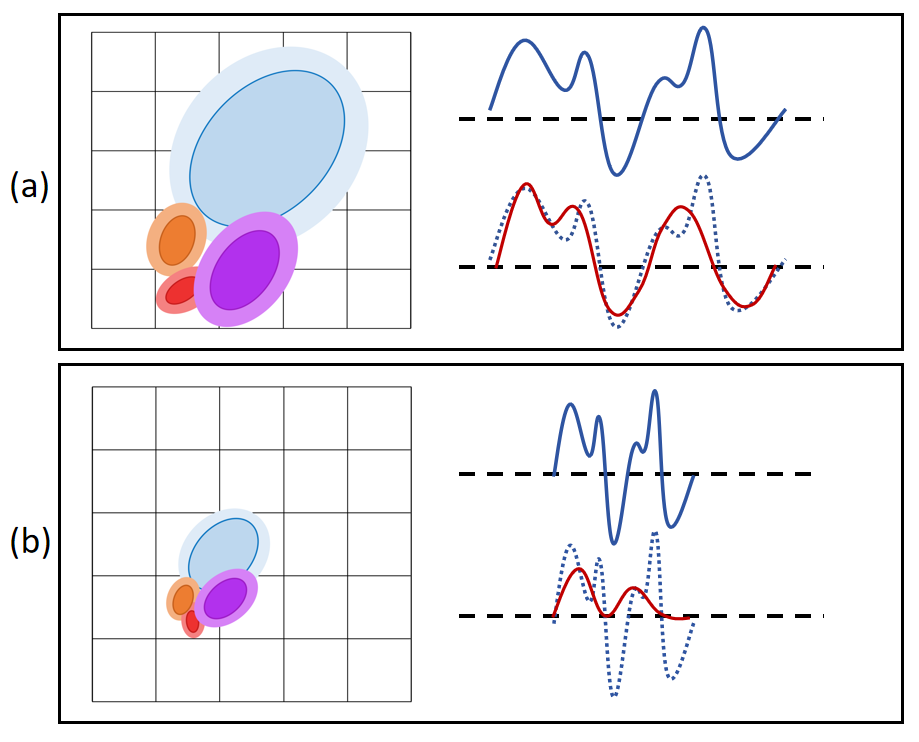}
% \vspace{-2em}
\caption{
\textbf{Signal visualization of the rendering process of Gaussian primitives.} (a) presents the zoomed-in view, while (b) displays the normal view. The left side is the Gaussians and the right side is their corresponding signals. The original signal is represented by the {\color{blue}blue curve}, and the sampled signal during rendering is indicated by the {\color{red}red curve}. After discrete sampling and filtering, high-frequency details—represented by the {\color{red}red Gaussian} and {\color{brown}brown Gaussian}—are observable only in the zoomed-in view but have negligible effects on the normal view, where high-frequency details are filtered out presented in the right bottom of (b).}
\label{fig:zoom_freq}
\centering
\end{figure}
%------------------------------------------

%------------------------------------------
\begin{figure*}[t!]
\centering
\includegraphics[width=0.9\textwidth]{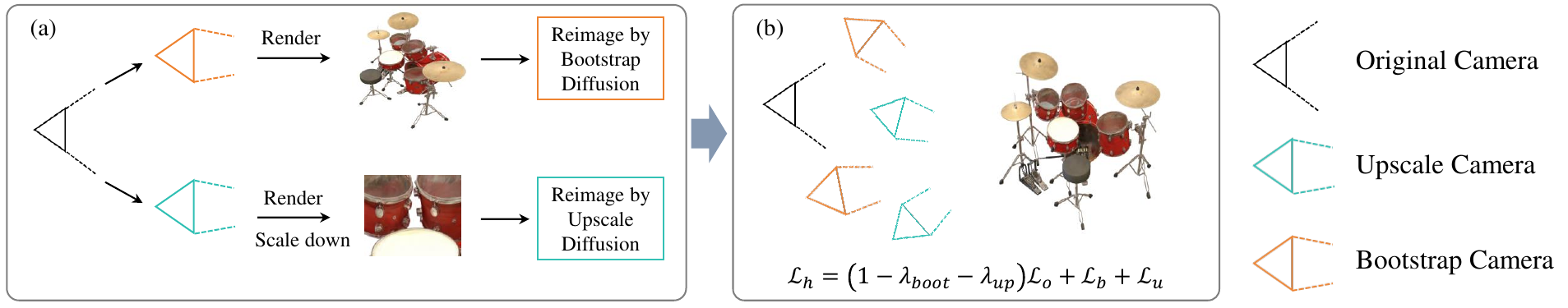}
% \vspace{-2em}
\caption{
\textbf{Visualization of our pipeline.} (a) Starting from an original camera, we first narrow its field of view to construct corresponding zoomed-in cameras and bootstrapping cameras. We render images from these cameras (at a reduced scale for upscaling). These new renderings are then fed into diffusion models for regeneration. (b) The regenerated images are used as pseudo ground truth to contribute to our hybrid loss function.}
\label{fig:pipeline}
\centering
\end{figure*}
%------------------------------------------

\subsection{Upscaling with Diffusion Models}

With the rapid advancement of diffusion models, their sampling procedures and design spaces have been extensively exploited in practical scenarios~\cite{karras2022elucidating, podell2023sdxl}. Our theoretical framework is primarily based on Latent Diffusion Models (LDMs)~\cite{rombach2022high}. Upscale diffusion models first encode an image into a latent space, and then modify and upsample the latent variables through the reverse diffusion process. Given the original pixel $p_o \in \RRR^{1\times 1 \times 3}$ and a zoomed-in tile $\mathbf{P} \in \RRR^{a\times a \times 3}$, $\mathbf{P}$ consists of a set of pixels with high-frequency details generated by upscaling diffusion models using an upscaling factor $a$. Based on Eq.~\ref{eq:image_intergation}, each pixel $p^{ij}$ in $\mathbf{P}$ can be represented:
\begin{equation}
\label{eq:upscal_diffusion}
p^{ij} = \sum_{t \in T} \boldsymbol{\epsilon}_\theta\left( \sqrt{\bar{\alpha}_t} \, \mathbf{x}_t(p_o, \, \sqrt{1 - \bar{\alpha}_t} \, \boldsymbol{\epsilon}) \right).
\end{equation}
We present the equation in this form for simplicity, as upsampling can be seamlessly integrated into the model's forward process.

High-frequency details in $\mathbf{P}$ originate from the random noise $\bepsilon$ introduced during each reverse diffusion process~\cite{rombach2022high}. While these details enhance the visual quality of the image, it is essential to ensure they perfectly align with the original lower-frequency ground truth $p_o$, as random noise can introduce significant deviations from the original content. Nevertheless, our thorough investigations reveal that the design space for high-frequency details, potentially generated by upscale diffusion models, is quite flexible. This flexibility stems from two primary factors: first, the rounding function applied after weighted average interpolation; and second, the application of a low-pass filter during the discrete signal reconstruction, particularly in the rendering process of 3D-GS.

\subsection{Interpolation Flexibility}
\label{sec:interpolation_flexibility}
To recover $p_o$ from $\mathbf{P}$ via interpolation, we perform weighted sampling followed by rounding the result to an integer within the range $[0, 255]$:
\begin{align}
\label{eq:down-interpolation}
p_o = &Round  ( \sum_{i}^{a}\sum_{j}^{a} \mathcal{W}(d^{ij}) p^{ij}  ), \\
where \quad d^{ij} &= \sqrt{\left(i-\frac{1}{2}a\right)^2 + \left(j-\frac{1}{2}a\right)^2}.
\end{align}
Here, $Round(\cdot)$ denotes the rounding function that maps values to their nearest integers within the specified range and $\mathcal{W}(\cdot)$ is a function to calculate the weight of pixels during interpolation. This rounding function introduces basic flexibility into the design space. If the absolute variation of the pixels in $\mathbf{P}$ compared to the pixel $p_o$ does not exceed a threshold $\tau_v$, such that the absolute variation after interpolation will not exceed 0.5, and the final result will remain unchanged. 

Since these variations are generally introduced by the noise $\bepsilon$ in Eq.~\ref{eq:upscal_diffusion} during each reverse diffusion step, we assume that given a noise threshold $\tau_{n}>0$, then for each $p^{ij}$ in $\mathbf{P}$, it holds that $\left|p_o - p^{ij}\right| \leq \tau_v$. This implies:
\begin{equation}
\label{eq:noise_limi}
\left| p_o - \sum_{t \in T} \boldsymbol{\epsilon}_\theta\left( \sqrt{\bar{\alpha}_t} \, \mathbf{x}_t(p_o, \,  \pm \sqrt{1 - \bar{\alpha}_t} \, \tau_{n}) \right) \right| \leq \tau_v.
\end{equation}
Thus, the problem is reduced to effectively regulating the normal noise $\bepsilon \sim \mathcal{N}(0, I)$ during the diffusion process.

We now focus on controlling the expectation of a set of independent normal distributions. Utilizing Chebyshev's inequality, we can express it in the following form:
\begin{theorem}
\label{theo: Chebyshev_ineq}
Mathematically, let \( X_1, X_2, \dots, X_n \) be i.i.d. random variables with expected value \( \mu = \mathbb{E}[X_i] \) and variance \( \sigma^2 = \operatorname{Var}(X_i) < \infty \), then, for any \( \varepsilon > 0 \):
\begin{align}
\mathbb{P}\left( \left| \frac{1}{n} \sum_{i=1}^{n} X_i - \mu \right| \geq \varepsilon \right) \leq \frac{\sigma^2}{n \varepsilon^2}.
\end{align}
\end{theorem}
As the sample size $n$ increases, the right-hand side $\frac{\sigma^2}{n \varepsilon^2}$ approaches zero, indicating that the sample mean converges in probability to the expected value $\mu$. And according to \textbf{Statistical Principles}:
\begin{condition}
\label{theo: statistic_states}
When the probability of an event occurring is below a certain threshold, it can be considered negligible for practical purposes in small samples.
\end{condition}
Combining these theories, it follows that for a given probability threshold $\tau_{p}$, there exists a sample size $n_s$ satisfying:
\begin{align}
\mathbb{P}\left( \left| \frac{1}{n_s} \sum_{i=1}^{n} \bepsilon \right| \geq \tau_{n} \right) \leq \tau_{p},
\end{align}
such that $\left| \frac{1}{n_s} \sum_{i=1}^{n} \bepsilon \right| \geq \tau_{n}$ is highly unlikely to occur during our sampling process, which we could consider to meet the requirements of the inequality in Eq.~\ref{eq:noise_limi} based on Condition~\ref{theo: statistic_states}.

In practice, we could sacrifice the ability and versatility of diffusion models but directly fix $\bepsilon$ with a small value less than $\tau_n$. However, based on our reasoning, we find that increasing the posterior sampling number could also achieve similar results because:
\begin{equation}
\label{eq:ineq_transformation}
\begin{aligned}
%&\left| p_o - \mathbb{E}[\sum_{t \in T} \bepsilon_\theta\left( \sqrt{\bar{\alpha}_t} \, \mathbf{x}_t(p_o, \, \sqrt{1 - \bar{\alpha}_t} \, \bepsilon) \right)]  \right|\\
& \left| \mathbb{E} \left[ \sum_{t \in T} \boldsymbol{\epsilon}_\theta\left( \sqrt{\bar{\alpha}_t} \, \mathbf{x}_t(p_o, \, \frac{1}{n_s}\sum_{i=1}^{n_s}\,\sqrt{1 - \bar{\alpha}_t} \, \bepsilon_i) \right) \right] \right|\\
& \geq \left| \frac{1}{n_s} \sum_{i}^{n_s} \mathbb{E} \left[ \sum_{t \in T} \boldsymbol{\epsilon}_\theta\left( \sqrt{\bar{\alpha}_t} \, \mathbf{x}_t(p_o, \, \sqrt{1 - \bar{\alpha}_t} \, \bepsilon_i) \right) \right] \right|. \\
\end{aligned}
\end{equation}
Therefore, similar to the approach used in Bootstrap-GS, increasing the number of diffusion samples can significantly reduce misalignment between generated fine-grained details and the original low-frequency ground truth. For a complete derivation and additional details, please refer to our \textbf{Supplementary Material}.

\subsection{Filter Application in 3D-GS}
\label{sec:filter_3DGS}
On the other hand, combining the 3D-GS rendering process, to render $p_o$ from the original viewpoint involves all high-frequency Gaussians visible in the zoomed-in tile $\mathbf{P}$. This inclusion will exceed the highest frequency permissible by the discrete sampling threshold described in Sec.~\ref{sec:img_distortion}. Therefore, a frequency filter $Fil(\cdot)$ is applied to the Gaussians visible in $\mathbf{P}$. The pixel $p_o$ rendered from the original view can then be approximated from the interpolation perspective as:
\begin{equation}
\label{eq:render-interpolation}
p_o = Round (\sum_{i}^{a}\sum_{j}^{a} \mathcal{W}(d^{ij}) Fil(p^{ij})  ).
\end{equation}
In 3D-GS~\cite{kerbl20233d, yu2023mipsplat, gao2024bootstrap3dgs}, this filter operates on small Gaussians that remain too diminutive to fill an entire pixel, even after extension. During the densification process of 3D-GS, high-frequency details are typically represented by smaller Gaussians. The filter $Fil$ not only acts as a barrier for very small Gaussians but also smooths the distribution of color density during rendering, as illustrated in Figure~\ref{fig:zoom_freq}. 

By integrating analyses from both the rendering and interpolation perspectives, we conclude that extremely small Gaussians aggregated within a single pixel have a negligible impact on rendering from relatively distant viewpoints. This negligible influence permits us to edit these Gaussians freely without affecting the overall image quality. Conversely, from an interpolation standpoint, the visible zoomed-in high-frequency Gaussians in distant views can still be stably and effectively modified while maintaining alignment with the ground truth of the original views by upscale diffusion models.

\subsection{Practical Implementation}
\label{sec:main_imple}
\boldparagraph{Loss Design} 
For each training camera $c_t$ and its corresponding ground-truth image $\mathbf{I}_t$, we adjust the translation scale of $c_t$ to obtain zoom-in cameras $\mathbf{c}_z$ using a set of scaling factors $\mathbf{a}$. We then upscale $\mathbf{I}_t$ using diffusion models with these scaling factors to generate zoom-in pseudo-ground-truth images $\mathbf{I}_z$. Our upscaling loss $\mathcal{L}_u$ is defined as the average $\mathcal{L}_1$ loss between the rendered images $\mathbf{I}_{r}$ under the zoom-in cameras $\mathbf{c}_z$ and their corresponding bootstrapped upscaled images $\mathbf{I}_z$, where $\mathcal{L}_{u} = \frac {\lambda_{\text{up}}}{M} \sum_{i}^{M} { \lVert \mathbf{I}_{r}^i - \mathbf{I}_z^i \lVert}$. Here $M$ is the total number of zoom-in scales and $\lambda_{\text{up}}$ is a loss scaling factor. Finally, we employ a hybrid loss $\mathcal{L}_h$ that combines the original 3D-GS loss $\mathcal{L}_o$, the bootstrapping loss $\mathcal{L}_b$ introduced in Sec.~\ref{sec:boot_intro} (if used), and the upscaling loss $\mathcal{L}_{u}$: $\mathcal{L}_h = (1-\lambda_{\text{boot}} - \lambda_{\text{up}})\mathcal{L}_o + \mathcal{L}_b + \mathcal{L}_{u}$.

\boldparagraph{Refinement of Bootstrapping Pipeline}
Despite its success, Bootstrap-GS has an unstable training process because of the variability of bootstrapped views. Our extensive experiments show that most fluctuations are not due to multi-view inconsistencies from bootstrapping but rather from the durability of bootstrapped views in early training stages. The rapid completion of 3D scenes means bootstrapped views before several hundred iterations are often largely misaligned with current views from a whole perspective. Additionally, some artifacts cannot be mitigated by merely expanding sampling views. The original 3D-GS generally lacks structural awareness due to its straightforward optimization. Even if artifacts in unseen views are modified, Bootstrap-GS may only adjust colors to appear more ``natural," but the underlying Gaussians—which ideally should not exist—remain. As a result, we not only limit the training duration of each bootstrapped view but also use the LOD structure proposed in Octree-GS \cite{ren2024octreegs}. For more details and experimental support, please refer to the \textbf{Supplementary Material}.

\boldparagraph{Cropped Upscaling and Bootstrapping}
Upscaling using diffusion models is a time-consuming process. To ensure proper alignment between the generated details and the original content, using extremely small diffusion steps is not acceptable. On top of that, in Bootstrap-GS, bootstrapping time is a significant concern, which may even triple the training time. Combining both upscaling and bootstrapping would render the training time prohibitive. Furthermore, while the color optimization in the original 3D-GS with spherical harmonics is relatively insensitive to bootstrapping alterations as they could be simply converted back independently, MLP-based color prediction methods like Scaffold-GS~\cite{lu2023scaffold} are highly susceptible to these changes. Altering the whole image, even with a small modification strength via diffusion models, can lead to a complete collapse in the training of the color MLP as it may contradict the previous training results. To address these issues, we have designed a cropping method. We find that reducing the input image size and the camera's field of view not only increases training speed but also improves the final performance for bootstrapping and upscaling. The whole pipeline is displayed in Figure~\ref{fig:pipeline}. For a detailed analysis, please refer to our \textbf{Supplementary Material}.

\section{Experiments}
\label{sec:exp}

\begin{table*}[htbp]
\centering
\renewcommand{\arraystretch}{1.15}
\setlength{\tabcolsep}{2pt}
\caption{Quantitative comparison on real-world datasets~\cite{barron2022mipnerf360,knapitsch2017tanks,hedman2018deep}. In our Stage 1 experiments, where we utilize the upscaling diffusion model, we observe not only an absence of performance degradation but also slight improvements in the evaluation metrics. In Stage 2, by combining upscaling with bootstrapping, we significantly enhance the overall performance in each scene while maintaining the increment of Gaussian counts in a negligible number. We have highlighted the \textbf{best} and \underline{second-best} results in each category.}
% \vspace{-6pt}
\label{tab:real_q}
\resizebox{1\linewidth}{!}{
\begin{tabular}{l|cccc|cccc|cccc}
\toprule
Dataset & \multicolumn{4}{c|}{Mip-NeRF360} & \multicolumn{4}{c|}{Tanks\&Temples} & \multicolumn{4}{c}{Deep Blending} \\
\begin{tabular}{c|c} Method & Metrics \end{tabular}  & PSNR\(\uparrow\) & SSIM\(\uparrow\) & LPIPS\(\downarrow\) & \#GS(k)/Mem & PSNR\(\uparrow\) & SSIM\(\uparrow\) & LPIPS\(\downarrow\) & \#GS(k)/Mem & PSNR\(\uparrow\) & SSIM\(\uparrow\) & LPIPS\(\downarrow\) & \#GS(k)/Mem \\
\midrule

Mip-NeRF360~\cite{barron2022mipnerf360} & 27.69 & 0.792 & 0.237 & - & 23.14 & 0.841 & 0.183 & - & 29.40 & 0.901 & 0.245 & - \\

3D-GS~\cite{kerbl20233d} & 27.54 & 0.815 & 0.216 & 937/786.7M & 23.91 & 0.852 & 0.172 & 765/430.1M & 29.46 & 0.903 & 0.242 & 398/705.6M \\

Mip-Splatting~\cite{yu2023mipsplat} & 27.61 & 0.816 & 0.215 & 1013/838.4M & 23.96 & 0.856 & 0.171 & 832/500.4M & 29.56 & 0.901 & 0.243 & 410/736.8M \\

Scaffold-GS~\cite{lu2023scaffold} & 27.90 & 0.815 & 0.220 & 666/197.5M & 24.48 & 0.864 & 0.156 & 626/167.5M & 30.28 & 0.909 & 0.239 & 207/125.5M \\

Boot-3D-GS~\cite{gao2024bootstrap3dgs} 
& \underline{28.32} & \underline{0.823} & \textbf{0.209} & 951/798.5M
& \underline{24.85} & 0.863 & 0.163 & 788/448.3M
& \textbf{31.43} & \underline{0.914} & \textbf{0.231} & 412/730.4M \\

Octree-GS~\cite{ren2024octreegs} & 28.05 & 0.819 & 0.214 & 657/\textbf{139.6M} & 24.68 & 0.866 & 0.153 & 443/\textbf{88.5M} & 30.49 & 0.912 & 0.241 & 112/\textbf{71.7M} \\

\hline

Octree-UP & 28.14 & 0.819 & 0.213 & 664/\underline{141.2M} & 24.78 & \underline{0.868} & \underline{0.152} & 463/\underline{92.5M} & 30.56 & \underline{0.914} & 0.239 & 114/\underline{73.1M} \\

Octree-UB & \textbf{28.53} & \textbf{0.825} & \underline{0.210} & 674/143.3M & \textbf{25.15} & \textbf{0.871} & \textbf{0.149} & 474/94.7M & \underline{31.23} & \textbf{0.919} & \underline{0.237} & 112/116.9M \\

\bottomrule
\end{tabular}}
\vspace{-1em}
\end{table*}

\subsection{Experimental Setup}
\paragraph{Dataset and Metrics.}
We conducted comprehensive evaluations on 29 scenes from public datasets, encompassing a diverse range of environments—from complex indoor settings and multi-scale scenarios to large urban landscapes. Specifically, our evaluation includes 9 scenes from Mip-NeRF360~\cite{barron2022mipnerf360}, 2 scenes from Tanks \& Temples~\cite{knapitsch2017tanks}, 2 scenes from DeepBlending~\cite{hedman2018deep}, 8 scenes from BungeeNeRF~\cite{xiangli2022bungeenerf}, and 8 objects from NeRF-Synthesis~\cite{martin2021nerf}.

Our primary focus is on comprehensively improving scene reconstruction without altering the structure of our frameworks. Therefore, we report only standard visual quality metrics such as PSNR, SSIM~\cite{wang2004image}, LPIPS~\cite{zhang2018unreasonable}, and the number of Gaussian primitives. For other relevant metrics like FPS, the variations compared to the original models remain minimal. The quantitative metrics averaged over all scenes and test frames are presented in the main paper, while detailed results for each scene are provided in the \textbf{Supplementary Material}.

\begin{figure*}[t]
	\centering
	\includegraphics[width=0.9\textwidth]{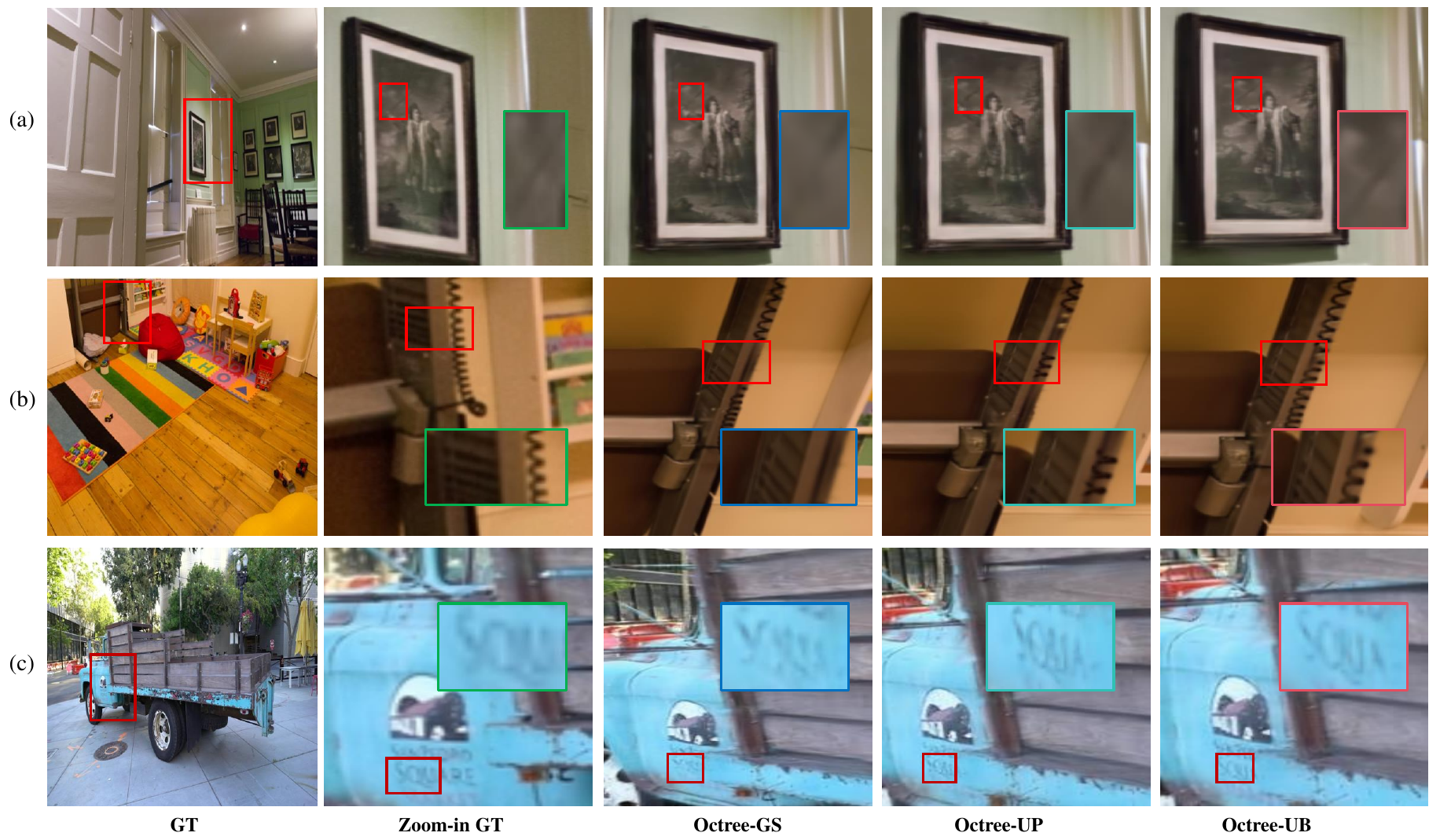}
	%\vspace{-10pt}
	\caption{\textbf{Main comparisons.} (a) For extremely small details present in the ground truth images, our method effectively completes and reconstructs these details. (b) In scenarios involving partially occluded views within the training datasets, our random upscale sampling technique enables the generation of fine-grained details that align with the ground truth. (c) For vague or indistinct details even in the ground truth, our upscaling diffusion models are capable of denoising the ground truth and generating high-frequency details. \textbf{Note}: All our renderings are produced from zoomed-in views, for which there is no directly aligned ground truth available.}
	\vspace{-10pt}
	\label{fig:main_comp}
\end{figure*}

\paragraph{Baseline and Implementation.}
We compare our method against the original 3D-GS~\cite{kerbl20233d}, Mip-Splatting~\cite{yu2023mipsplat}, Bootstrap-GS~\cite{gao2024bootstrap3dgs}, the original Scaffold-GS~\cite{lu2023scaffold}, and Octree-GS~\cite{ren2024octreegs}. We also report the results of MipNeRF360~\cite{barron2022mipnerf360} for rendering quality comparisons. To ensure fair comparisons with the original results, we kept all modeling parameters unchanged and only incorporated bootstrapping and upscaling techniques. Additionally, as Octree-GS \cite{ren2024octreegs} has been implemented across various frameworks, including 3D-GS and Scaffold-GS \cite{lu2023scaffold}, we focus primarily on its most successful version, the Scaffold-GS implementation. In the following sections, all references to Octree-GS default to its implementation on Scaffold-GS. 
%For more details on Octree-GS, please refer to~\cite{ren2024octreegs}.

\begin{figure*}[t]
	\centering
	\includegraphics[width=0.95\linewidth]{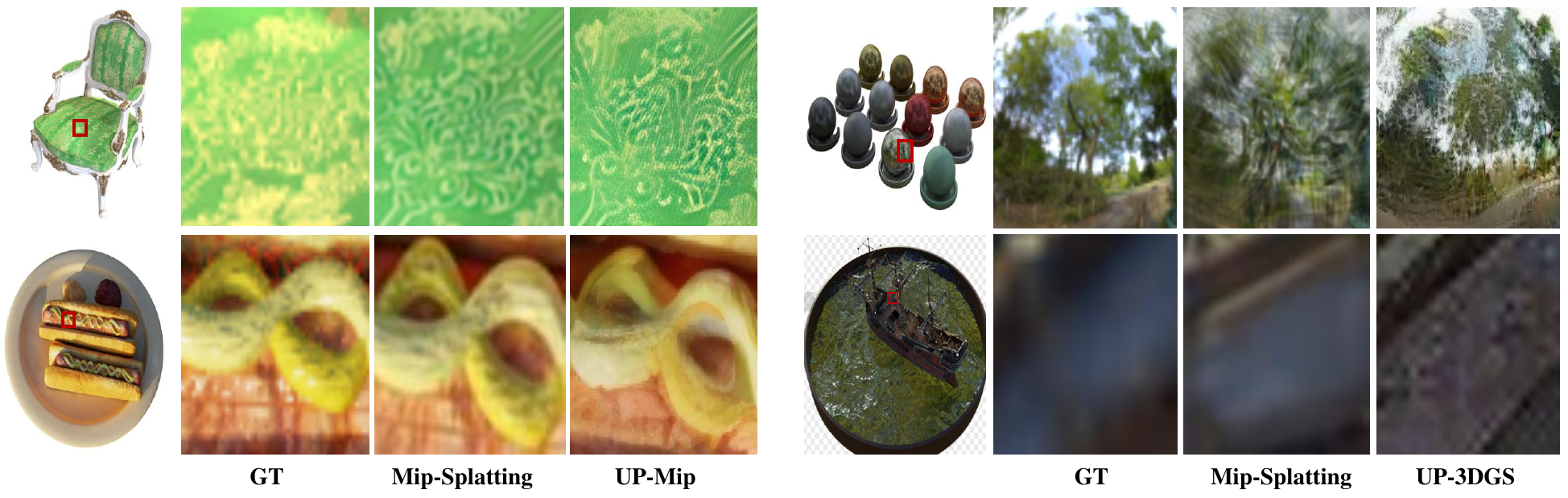}
	%\vspace{-10pt}
	\caption{\textbf{Object rendering comparisons.} Our method enables the generation of fine-grained, flexible, and scene-aligned details using upscaling diffusion models on zoomed-in scales without impacting the integrity of the original rendering on normal scales.}
	\vspace{-10pt}
	\label{fig:zoom_obj}
\end{figure*}

Due to the distinct characteristics of each dataset, we apply task-specific training schedules. For foundational outdoor and indoor datasets, Mip-NeRF360, Tanks\& Temples, and DeepBlending, we conduct two-stage training per scene using the Octree-GS framework. In the first stage, we incorporate upscaling with a factor of $a = 2$, denoted as \textbf{Octree-UP}, leveraging the open-source diffusion model \textbf{SD-X2-Latent-Upscaler}~\cite{rombach2022high} to generate zoom-in details, which are reintegrated into training. In the second stage, in addition to upscaling, we integrate bootstrapping using \textbf{Stable Diffusion SDXL-Turbo} along with our refined bootstrapping pipeline, denoted as \textbf{Octree-UB}. During this stage, we fine-tune \textbf{SDXL-Turbo} based on~\cite{gao2024bootstrap3dgs} for each scene while keeping the scaling factor unchanged. %Given that the Octree-GS version we use here relies on MLP-based color prediction, which is sensitive to bootstrapping disturbances as described in Sec.~\ref{sec:main_imple}, we also train an additional version using Octree-3D-GS as our framework—an implementation of Octree-GS based on the original 3D-GS structure.
Given that the Octree-GS version we use here, which relies on MLP-based color prediction, is sensitive to bootstrapping disturbances as described in Section \ref{sec:main_imple}, we also train an additional version using Octree-3D-GS as our framework. This framework is an implementation of Octree-GS based on the original 3D-GS structure.

For the multi-scale dataset BungeeNeRF, we apply only upscaling diffusion models on Octree-GS, conducting two-stage training with different scaling factors. Specifically, in the first stage, we use a scaling factor of \( a = 2 \), denoted as \textbf{Octree-UPx2}, and in the second stage, we apply scaling factors \( \mathbf{a} = \{2, 4\} \), denoted as \textbf{Octree-UPx4}. For NeRF-Synthesis, we apply upscaling diffusion models with scaling factors \( \mathbf{a} = \{4, 8\} \) on the original 3D-GS (\textbf{UP-3DGS}) and Mip-Splatting (\textbf{UP-Mip}) frameworks, as the LOD structure is ineffective in this context, and MLP-based color prediction tends to produce lower-frequency details in highly zoomed-in views. The upscaling diffusion models used are \textbf{SD-X2-Latent-Upscaler} for lower scaling factors and \textbf{SD-X4-Upscaler} for higher scaling factors across each dataset. For further implementation details and explanations, please refer to our \textbf{Supplementary Material}.

\begin{table*}[htbp]
% \vspace{6pt}
\centering
\renewcommand{\arraystretch}{1.15}
\setlength{\tabcolsep}{3pt}
\caption{
Quantitative comparison on the BungeeNeRF~\cite{xiangli2022bungeenerf} dataset. 
% Considering the dataset's multi-scale nature, w
We present evaluation metrics for each scale and their averages across all four scales. Scale-1 corresponds to the closest viewpoints, while Scale-4 encompasses the entire landscape. Our method shows increasing effectiveness as the scale increases, all while maintaining a minimal increase in the number of Gaussians.}
% \vspace{-6pt}
\label{tab:bungeenerf}
\resizebox{1\linewidth}{!}{
\begin{tabular}{l|cccc|cc|cc|cc|cc}
\toprule
~~~Dataset & \multicolumn{4}{c|}{BungeeNeRF (Average)} & \multicolumn{2}{c|}{Scale-1} & \multicolumn{2}{c|}{Scale-2} & \multicolumn{2}{c|}{Scale-3} & \multicolumn{2}{c}{Scale-4} \\

\begin{tabular}{c|c} Method & Metrics \end{tabular}  & PSNR\(\uparrow\) & SSIM\(\uparrow\) & LPIPS\(\downarrow\) & \#GS(k)/Mem & PSNR\(\uparrow\)  & \#GS(k) & PSNR\(\uparrow\) & \#GS(k) & PSNR\(\uparrow\) & \#GS(k) & PSNR\(\uparrow\) & \#GS(k) \\
\midrule

3D-GS~\cite{kerbl20233d} & 27.79 & 0.917 & 0.093 & 2686/1792.3M & 30.00 & 522 & 28.97 & 1272 & 26.19 & 4407 & 24.20 & 5821 \\

Mip-Splatting~\cite{yu2023mipsplat} & 28.14 & 0.918 & 0.094 & 2502/1610.2M & 29.79 & 503 & 29.37 & 1231 & \underline{26.74} & 4075 & 24.44 & 5298 \\

Scaffold-GS~\cite{lu2023scaffold} & 28.16 & 0.917 & 0.095 & 1652/319.2M & 30.48 & 303 & 29.18 & 768 & 26.56 & 2708 & 24.95 & 3876 \\

Octree-GS~\cite{ren2024octreegs} & 28.39 & \underline{0.923} & \underline{0.088} & 1474/\textbf{296.7M} & \underline{31.11} & 486 & 29.59 & 1010 & 26.51 & 2206 & 25.07 & 2167 \\

\hline

Octree-UPx2 & \underline{28.43} & \underline{0.923} & \underline{0.088} & 1521/\underline{306.2M} & \textbf{31.14} & 490 & \underline{29.63} & 1022 & 26.83 & 2234 & \underline{25.28} & 2191 \\

Octree-UPx4 & \textbf{28.48} & \textbf{0.924} & \textbf{0.087} & 1568/315.6M & 31.08 & 493 & \textbf{29.72} & 1031 & \textbf{26.75} & 2249 & \textbf{25.41} & 2213 \\

\bottomrule
\end{tabular}}
% \vspace{-1.4em}
\end{table*}

\subsection{Results Analysis}

In both outdoor and indoor scene results, as shown in Table \ref{tab:real_q}, Figures~\ref{fig:main_comp} and \ref{fig:dilation} (left), our first-stage experiment, \textbf{Octree-UP}, effectively preserves original content while enhancing fine details that are otherwise obscured in the ground truth. In the second stage, \textbf{Octree-UB}, our refined bootstrapping pipeline builds on \textbf{Octree-UP} to achieve substantial metric improvements with moderate added training time (see Sec.~\ref{sec:further_detail}). Additionally, training within the \textbf{Octree-3D-GS} framework produces further gains, significantly outperforming the original Bootstrap-GS, as detailed in our \textbf{Supplementary Material}.

For the multi-scale datasets shown in Table~\ref{tab:bungeenerf}, we observe that in most scenarios, fine-grained details generated by upscaling diffusion models are largely imperceptible in the test data, while only visible in zoomed-in views as shown in Figure ~\ref{fig:zoom_artifacts}. Consequently, modifications mainly impact the overall structure in training, resulting in only minor metric improvements. However, as scale increases, the benefits of our upscaling method become clearer, with structural adjustments to the scenes contributing more substantially to performance gains.

Finally, for object training, each framework offers unique advantages, as shown in Figure \ref{fig:zoom_obj}. Under \textbf{UP-Mip}, with the application of a 3D smoothing filter and a 2D Mip filter, high-frequency details appear smoother. In contrast, \textbf{UP-3DGS} produces sharper, though occasionally more variable, details. Both methods demonstrate the ability to generate high-frequency details out of training data compared with blurry ground truth, significantly enriching 3D reconstruction results. Further analyses are in our \textbf{Supplementary Material}.

\subsection{Further Details}
\label{sec:further_detail}

\begin{figure}[t!]
\centering
\includegraphics[width=0.475\textwidth]{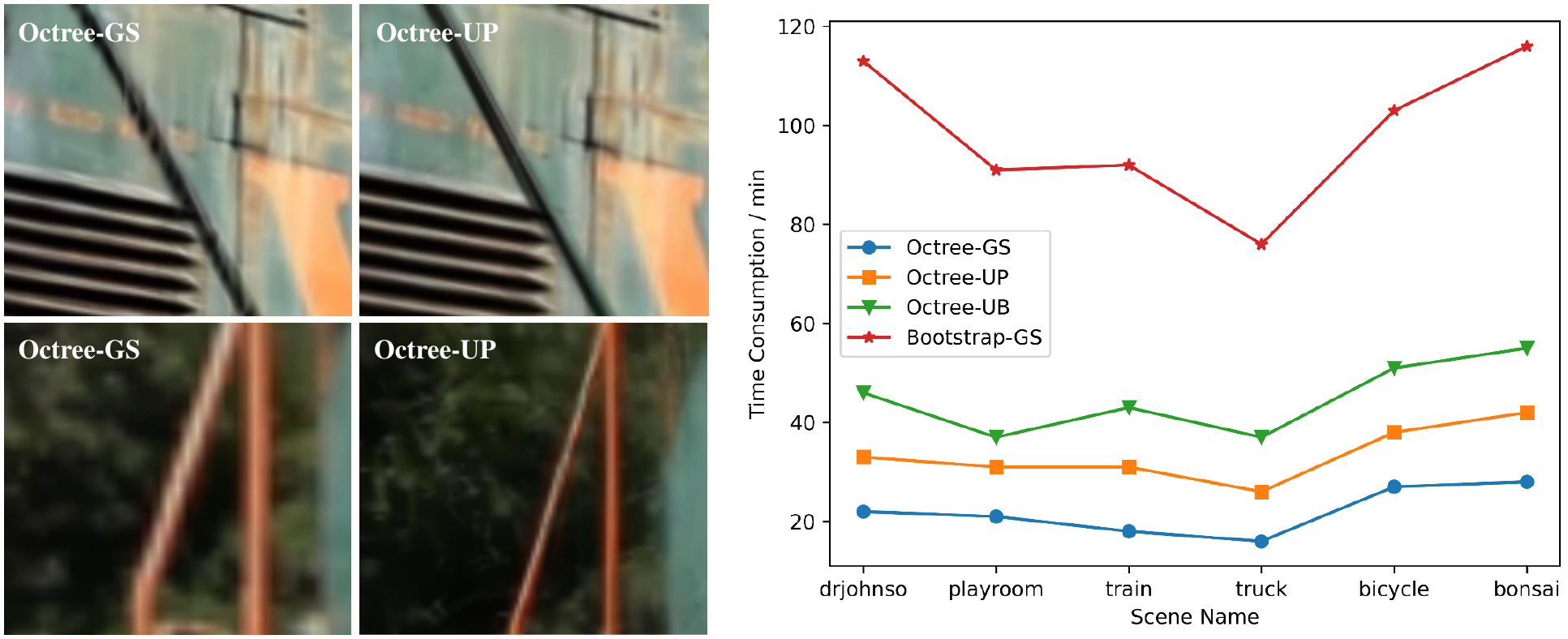}
% \vspace{-2em}
\caption{Left: Visualization of zoomed-in $\times$4 views between original Octree-GS and our Octree-UP in \textbf{Train}~\cite{knapitsch2017tanks}. Our renderings exhibit less dilation and sharper details. Right: Time comparisons among different methods. Our Octree-UP and Octree-UB achieve significant progress on performance and largely reduce the time consumption compared with Bootstrap-GS.
}
\label{fig:dilation}
\centering
\end{figure}

\boldparagraph{Time Consumption}
Our training process is significantly faster than the original bootstrapping pipeline, as shown in Figure \ref{fig:dilation} (right). With the upscale diffusion model, \textbf{Octree-UP} requires only 10 additional minutes for outdoor and indoor scenes. Even with bootstrapping integration, our training time of \textbf{Octree-UB} averages under one hour per scene. For the multi-scale training in BungeeNeRF datasets \cite{xiangli2022bungeenerf}, we only upscale part of the training data, which further accelerates the training process. Additionally, We also trained the original model for the same length of time for fair comparisons. The results indicate that extended training contributes minimal improvements in metrics and does not further alleviate zoomed-in artifacts, but even \textbf{worsens}. For more details please refer to our \textbf{Supplementary Material}.

\boldparagraph{Flexibility of Zoom-in Details}
Our experiments reveal that in most scenes small artifacts like ``partial breakage" and minor distortion begin to appear when zooming in 2$\times$ on the original camera views. When zooming in 4$\times$, some regions exhibit complete breakdown. This suggests that the more a reconstructed scene collapses in zoomed-in views, the more flexibility we have to use upscale diffusion models. In these views, conditional upscaling generation can even be used to precisely design the desired details. For instance, in Figure~\ref{fig:zoom_obj}, our generation may slightly deviate from the original content, yet the evaluation metrics in standard views remain comparable to original frameworks.

\boldparagraph{Control of Zoom-in Details}
To control the flexibility of zoom-in details, we could finetune the upscale diffusion models to achieve more aligned generation. Additionally, we could apply a gradual zoom-in scaling approach, for instance, using a set of scales such as \( \{2, 4, 6, 8\} \) rather than our \( \{4, 8\} \) for object reconstruction, to progressively refine details at each zoom level. The effectiveness of this approach could be validated through the application of BungeeNeRF in Figure~\ref{fig:zoom_artifacts}, where generated details closely align with the original content. However, in our paper, we opted not to implement this gradual approach because we aimed to demonstrate the flexibility of the design space. Instead, we present it as a solution for achieving higher fidelity in practical implementations.

\iffalse
\begin{figure}[t!]
\centering
\includegraphics[width=0.35\textwidth]{figs/time_comp.png}
% \vspace{-2em}
\caption{Time comparisons among different methods. Our Octree-UP and Octree-UB achieve significant progress on performance and largely reduce the time consumption compared with Bootstrap-GS.
}
\label{fig:train_time}
\centering
\end{figure}
\fi
\section{Conclusion}
We introduce a new training pipeline for 3D-GS that integrates both Bootstrap-GS and upscaling diffusion models. Our approach includes a comprehensive theoretical analysis and has demonstrated remarkable practical success. It significantly reduces zoom-in artifacts commonly seen in diverse 3D-GS frameworks and leverages a flexible design space for detailed zoom-in elements, enhancing the expressive quality of 3D reconstruction. Additionally, our refined bootstrapping and upscaling pipelines result in a much faster training process compared to the original Bootstrap-GS. Our method achieves state-of-the-art results with improved generalization to out-of-distribution camera poses, all while adding minimal extra training time.

{
    \small
    \bibliographystyle{ieeenat_fullname}
    \bibliography{main}
}
\clearpage

\newpage
\section{Supplementary Material}
\noindent \textbf{Overview.} This supplementary is structured as follows:
(1) The first section elaborates on our further analyses and implementation details, (2) and additional experimental results are also presented.

\noindent \textbf{Erratum.} We have identified an error in Table~\ref{tab:real_q}, which presents the main outdoor and indoor comparison. Specifically, the values for \textbf{\#GS(k)/Mem} associated with \textbf{Deep Blending} are incorrect. The correct values should be \textbf{117/75.6M}.

\subsection{3D Gaussian Splatting}
\label{sec:gs_intro}
The geometry of each scaled 3D Gaussian starts with a set of points derived from Structure-from-Motion (SfM), each point is designated as the position (mean) $\mu$ of a 3D Gaussian: $G(x) = e^{-\frac{1}{2} (x-\mu)^T \Sigma^{-1} (x-\mu)}.$
\iffalse
\begin{equation}
G(x) = e^{-\frac{1}{2} (x-\mu)^T \Sigma^{-1} (x-\mu)},
\label{gs}
\end{equation}
\fi
To constrain $\bm{\Sigma}$ to the space of valid covariance matrices, a semi-definite parameterization is used. Then, the covariance matrix $\bm{\Sigma}$ is constructed utilizing a scaling matrix $\bm{S}$ and a rotation matrix $\bm{R}$, ensuring that it remains positive semi-definite: $\bm{\Sigma} = \bm{RS}\bm{S}^T\bm{R}^T.$

To render an image for a given viewpoint efficiently, 3D-GS uses tile-based rasterization. This process involves initially converting the 3D Gaussians $G(x)$ into 2D Gaussians $G'(x)$ on the image plane as a result of the projection process described in \citep{zwicker2001ewa}. Subsequently, a specially designed tile-based rasterizer sorts these 2D Gaussians and applies $\alpha$-blending: 
%\begin{footnotesize}
\begin{equation}
C(x') = \sum_{i \in N} c_i \sigma_i \prod_{j=1}^{i-1} (1 - \sigma_j), \quad \sigma_i = \alpha_i G'_{i}(x') ,
\label{gs_rendering}
\end{equation}
%\end{footnotesize}
where $x'$ is the queried pixel position and $N$ denotes the number of sorted 2D Gaussians associated with the queried pixel.

\subsection{Theoretical Reasoning Completion}
In this section, we aim to complete the logical chain outlined in Sec.~\ref{sec:interpolation_flexibility}. We begin by examining the left-hand side of the inequality in Eq. \ref{eq:ineq_transformation}, which is given by:
\begin{equation}
\label{eq:left_term_eq}
\left| \mathbb{E} \left[ \sum_{t \in T} \boldsymbol{\epsilon}_\theta\left( \sqrt{\bar{\alpha}_t} \, \mathbf{x}_t(p_o, \, \frac{1}{n_s}\sum_{i=1}^{n_s}\,\sqrt{1 - \bar{\alpha}_t} \, \bepsilon_i) \right) \right] \right|.
\end{equation}
As discussed in Sec.~\ref{sec:interpolation_flexibility}, this term can be interpreted as a \textbf{a global sample with an averaged collection of $n_s$ noise terms for each diffusion time step}. Each averaged collection of noise terms attains the maximum expected value satisfying the inequality in Eq.~\ref{eq:noise_limi} based on Theorem~\ref{theo: statistic_states}. Conversely, the right-hand side of the inequality in Eq.~\ref{eq:ineq_transformation} is expressed as:
\begin{equation}
\label{eq:right_term_eq}
\left| \frac{1}{n_s} \sum_{i}^{n_s} \mathbb{E} \left[ \sum_{t \in T} \boldsymbol{\epsilon}_\theta\left( \sqrt{\bar{\alpha}_t} \, \mathbf{x}_t(p_o, \, \sqrt{1 - \bar{\alpha}_t} \, \bepsilon_i) \right) \right] \right|.
\end{equation}
This expression can be viewed as \textbf{averaged $n_s$ global samples, each with a single noise term for each diffusion time step}. 

To simplify the analysis, we omit the term $p_o$ in the function $\mathbf{x}_t$ since it remains constant in both Eq.~\ref{eq:left_term_eq} and Eq.~\ref{eq:right_term_eq}, and there is no significant dependence between $p_o$ and $\bepsilon$. Consequently, $\mathbf{x}_t$ becomes a function solely of $\bepsilon$. And because $\bar{\alpha}_t$ is a pre-defined constant, we can merge $\bepsilon_\theta\left( \sqrt{\bar{\alpha}_t} \, \mathbf{x}_t(\sqrt{1 - \bar{\alpha}_t} \, \bepsilon_i) \right)$ into a simple function $\mathbf{f}_t(\bepsilon)$. By applying \textbf{Jensen's Inequality}, we reduce the problem to examining the convexity or concavity of the function $\mathbf{f}_t(\bepsilon)$ for each diffusion time step $t$. 

However, since $\boldsymbol{\epsilon}_\theta$ is a neural network comprising convolutions and nonlinear activation functions, it does not exhibit clear convexity or concavity properties. Therefore, we shift our focus to the behavior of its latent space to approximate the function's response. Specifically, we investigate whether large variations in the sampled noise $\bepsilon$ lead to significant fluctuations in the output.

To address this, we consider the properties of diffusion models as defined in~\cite{ho2020denoising,nichol2021improved,song2020denoising}. During the reverse diffusion process, the term $\sqrt{1 - \bar{\alpha}_t}$ diminishes progressively to zero at an exponential rate due to the cumulative product. This allows us to neglect time steps that contribute minimal variations and only focus on the initial reverse diffusion steps, which have a significant impact on the model's output.

Furthermore, based on the foundational theory of nonequilibrium thermodynamics in diffusion models~\cite{sohl2015deep}, a well-trained diffusion model is capable of recovering the original image from completely random noise after infinite reverse diffusion steps. This implies that initial noise fluctuations have a negligible effect on the final output. The model inherently possesses robustness, constraining large noise perturbations introduced in earlier time steps within an acceptable range.

Additionally, as demonstrated in DCSN~\cite{song2020score}, to address distribution challenges in low-dimensional spaces, noise is added to the original data during training to smooth the distribution, yet diffusion models still achieve successful results. This indicates that the latent space of diffusion models is sufficiently expressive such that local noise with significant variations in earlier time steps does not disrupt the expected distribution in the final output.

Therefore, we can reasonably assume that $\mathbf{f}_t$ behaves as a concave function satisfying the inequality $\mathbf{f}_t(\mathbb{E}[\bepsilon]) \geq \mathbb{E} \left[ \mathbf{f}_t(\bepsilon) \right]$. Under this assumption, the inequality in Eq.~\ref{eq:ineq_transformation} holds true, completing the logical chain of our argument.

\subsection{Analysis of Bootstrapping Deficiency}
This section aims to provide explanations for Sec.~\ref{sec:main_imple}. As outlined in Bootstrap-GS~\cite{gao2024bootstrap3dgs}, Gaussians with higher variations in gradient direction after bootstrapping tend to generate relatively consistent lower-frequency details due to the average bootstrapping loss. This enhancement improves evaluation metrics without causing significant fluctuations during training. On the other hand, the majority of out-of-view Gaussians—those with small sizes and fine-grained details—are generally unseen or occluded during rendering in test views, thanks to the filtering process introduced in Sec.~\ref{sec:filter_3DGS}. These Gaussians are essentially undetectable unless one explores novel-view renderings far beyond the training and testing sets. Consequently, bootstrapping effectively serves its intended purpose without introducing large training fluctuations.

However, the issue of large training fluctuation arises from the mismatch between the bootstrapping frequency and the rate of scene integration during training. Specifically, in the early stages of training, the reconstruction progresses so rapidly that after several hundred iterations, the current views represented by the trained Gaussian model differ significantly from those rendered a few hundred iterations earlier. Bootstrapping these misaligned views further introduces distortion. Consequently, during the early training phase, the bootstrapped views are significantly misaligned with the overall context and contribute minimally to unseen views after subsequent iterations. This misalignment can even lead to the degradation of the entire scene reconstruction.

On the other hand, the Gaussian model proposed in 3D Gaussian Splatting (3D-GS)~\cite{kerbl20233d} employs a straightforward optimization process. During reconstruction, when misalignments occur, the Gaussian model disperses Gaussian primitives to fill those regions with visually consistent colors. However, due to the incompleteness of scenes and a deficiency in training sampling, some excessively dispersed Gaussian primitives remain unmodified. Even with the application of bootstrapping, these Gaussians can easily adjust their Spherical Harmonics to compensate for misaligned areas without reducing their opacities. Consequently, Gaussian primitives that ideally should not exist persist throughout the reconstruction.

\subsection{Cropped Bootstrapping}
For bootstrapping, we proactively narrow the field of view of the novel-view camera to limit the size of the rendered image by a scale factor $s_{nar}$. In our experiments, we set $s_{nar} = 0.4$. As a result, the bootstrapped image becomes $\frac{4}{25}$ of its original size, which largely reduces the bootstrapping time. Moreover, our experiments indicate that most parts of the bootstrapped images exhibit no strong artifacts. While the color optimization in the original 3D-GS with spherical harmonics is relatively insensitive to bootstrapping alterations as they could be simply converted back independently, MLP-based color prediction methods like Scaffold-GS~\cite{lu2023scaffold} are highly sensitive to changes in the entire image. Altering the whole image, even with a small modification strength via diffusion models, can lead to a complete collapse in the training of the color MLP as it may contradict the previous training results. Therefore, cropping the bootstrapped image is conducive for stable training.

\subsection{Cropped Upscaling}
While the details generated by diffusion models during upscaling are generally consistent with the original content, the generation process is computationally intensive. Furthermore, zoomed-in views often exhibit noticeable artifacts. If we employ upscaling diffusion models to amplify these artifacts, the results may become unstable. To address this issue, we reduce both the field of view of the zoomed-in camera used for upscaling and scale down the rendering size by a factor of $a$. For example, with an upscale factor $a=2$ and the narrow factor $s_{nar}=0.5$, we render the image at $\frac{1}{4}$ of its original height and width, resulting in a rendered image that is $\frac{1}{16}$ of its original size. Our approach is based on the same principle described in Sec.~\ref{sec:filter_3DGS}: enlarging the pixel size can effectively filter out zoomed-in small Gaussian primitives. Zoom-in artifacts occur only when sizes of Gaussian primitives are large enough to span at least one pixel. By reducing the rendering size, we use a smaller image to encompass more content. Therefore, further decreasing the rendering size is analogous to the effects depicted in Fig.~\ref{fig:zoom_freq}(b), where the pixel size is extended. Subsequently, after applying upscaling, we obtain a more faithful rendering.

\subsection{Implementation details}
\label{sec:implementaion_details}

Our experimental setup for each framework model is consistent with that of Octree-GS~\cite{ren2024octreegs}. All scenes are trained on NVIDIA RTX 4090 GPUs.

%\paragraph{Basic Configurations.}
\subsubsection{Basic Configurations}
Following the methodology of Octree-GS~\cite{ren2024octreegs}, we trained all frameworks on basic indoor, outdoor, and multi-scale datasets for 40,000 iterations. In the Octree-Scaffold-GS framework, densification begins at iteration 500, while in the Octree-3D-GS framework, it starts at iteration 1,500. For both frameworks, the densification process concludes at iteration 25,000. For object training, we trained both Mip-Splatting and 3D-GS for 30,000 iterations. In these cases, densification commences at iteration 1,500 and ends at iteration 15,000. All other configurations, such as the learning rate and loss rate schedule, remained unchanged from their original frameworks.

%------------------------------------------
%\vspace{-2em}
\begin{table*}[htbp]
\centering
\renewcommand{\arraystretch}{1.15}
\setlength{\tabcolsep}{1pt}
\caption{Results for Tanks\&Temples \cite{knapitsch2017tanks} with different training time. Here, we denote Octree-Scaffold-GS as Octree-GS for simplicity. And for iteration counts, each number refers to $\times$10,000 iterations.}
%\vspace{-6pt}
%\resizebox{\linewidth}{!}
{
%\footnotesize
\small
%\begin{tabularx}{0.8\textwidth}{l|ccc|ccc}
\begin{tabular}{l|ccc|ccc}
\toprule
Dataset & \multicolumn{3}{c|}{Truck} & \multicolumn{3}{c}{Train} \\
\begin{tabular}{c|c}
\label{tab:time_comp_tnt}
Method & Metrics ~~\end{tabular}  & ~PSNR $(\uparrow)$ ~ & SSIM $(\uparrow)$ ~ & LPIPS $(\downarrow)$ ~ & ~ PSNR $(\uparrow)$ ~ & SSIM $(\uparrow)$ ~ & LPIPS $(\downarrow)$  \\
\midrule
Octree-GS-4 \cite{ren2024octreegs} & 26.24 & 0.894 & 0.122  & 23.11 & 0.838 & 0.184  \\
Octree-GS-5  & 26.21 & 0.893 & 0.122  & 23.07 & 0.837 & 0.185  \\
Octree-GS-6 & 26.22 & 0.894 & 0.122  & 23.08 & 0.837 & 0.184  \\
Octree-GS-7  & 26.22 & 0.893 & 0.122  & 23.07 & 0.837 & 0.185  \\
Octree-GS-8  & 26.24 & 0.894 & 0.122  & 23.12 & 0.838 & 0.184  \\
\midrule
Octree-UB & \textbf{26.47} & \textbf{0.896} & \textbf{0.120}  & \textbf{23.83} & \textbf{0.846} & \textbf{0.179}  \\
\bottomrule
\end{tabular}
}
\end{table*}
%------------------------------------------

%\paragraph{Bootstrapping Configurations.} 
\subsubsection{Bootstrapping Configurations}
Compared to Bootstrap-GS~\cite{gao2024bootstrap3dgs}, we increased the frequency of bootstrapping but delayed its implementation in our experiments. Bootstrapping was initiated at iteration 20,000 and concluded at iteration 38,000. Specifically, we performed bootstrapping every 2,000 iterations. During each bootstrapping interval, we conducted 750 iterations. This application ensures bootstrapped views are properly integrated and aligned with the ongoing training of the scene. 

For each training camera, we construct 2 of its randomly generated cameras and put them back to training the same as in Bootstrap-GS. For random cameras, we altered both the rotation matrices $\bm{R}$ and the translation vectors $\bm{t}$ by adding random noise with scaling factors of 0.2 and 0.1, respectively (after which $\bm{R}$ was re-normalized to ensure it remained a valid rotation matrix). Then we use a narrowing factor of 0.4 to scale down the field of view of bootstrapped cameras, with its rendered size reduced accordingly. Additionally, with structural improvement, the implementation of upscale diffusion models in Bootstrap-GS~\cite{gao2024bootstrap3dgs} becomes no longer unnecessary, so we discard such integration.

The implementation of bootstrapping diffusion models is relatively flexible, and customized diffusion models can also be effective. We utilize the text prompt \textbf{``sharp, denoise, original content, natural, detailed, 8k"} to guide conditional generation, employing a scaling factor of $1.0$. Based on prior investigations~\cite{gao2024bootstrap3dgs} and our experimental results, we find that as long as the sampling number is sufficiently large, minor adjustments in the configurations of diffusion models offer negligible improvements. Furthermore, the influence of the inference time step is minimal. In our experiments, we employ a linearly decaying guidance strength ranging from $0.06$ to $0.01$. For each value of the guidance strength, we dynamically adjust the total number of inference steps but perform only a single inference step during training.

%\paragraph{Upscaling Configurations.}
\subsubsection{Upscaling Configurations}
The upscaled cameras are derived directly from the training cameras by adjusting their scale to create zoomed-in views. Specifically, for each training camera, we modify its scale parameter to obtain a corresponding zoomed-in camera. Similar to the bootstrapping process, we add 2 additional random cameras for each training camera, altering only their rotation parameters randomly while keeping the scaling factor constant across all expanded zoomed-in cameras. Additionally, for multi-scale datasets BungeeNeRF \cite{xiangli2022bungeenerf}, we only apply our upscaling approach on the first 100 training cameras, as these cameras are gradually zoomed out, which further reduces the training time.

%the $22,000^{th}$ iteration
We initiate the upscaling process at iteration 22,000 and conclude it at iteration 38,000, using an upscaling interval of 2,000 iterations. Unlike bootstrapping, which can introduce training disturbances, upscaled regeneration remains relatively faithful to the original data. Therefore, during each interval, we incorporate these upscaled cameras for 1,000 iterations of training, leaving the remaining 1000 iterations unchanged. The narrowing factor is set to 0.5, effectively reducing the field of view by half, which is also sufficient to maintain training efficiency. 

In configuring the upscaling diffusion model, we fix the inference time step at 15 for all experiments. As demonstrated in Sec.~\ref{sec:interpolation_flexibility} and Sec. \ref{sec:filter_3DGS}, the design space permits flexibility for fine-grained details, so we employ the conditional generation with prompt \textbf{``sharp, detailed, denoise, 8k"}. To ensure the alignment of details generated by the upscaling diffusion models, we set the noise generator seed to a small value of $22$, and the conditional guidance for upscaling is merely 1.0.

%\paragraph{Loss Weights.} 
\subsubsection{Loss Weights}
We assign different values to the bootstrapping loss weight $\lambda_b$ and the upscaling loss weight $\lambda_u$ in Sec.~\ref{sec:main_imple} due to scene alignment issues discussed in Section~\ref{sec:main_imple}. Specifically, we set \( \lambda_b = [0.15, 0.1] \) and \( \lambda_u = [0.1, 0.05] \). Both of these factors are very small because of the disturbance issue on color prediction MLP. We uniformly divide the training duration of bootstrapping and upscaling in each interval and apply different scales to their respective losses accordingly. Additionally, the upscale loss could be amplified and directly added to the training loss without the scaling factor $\lambda_u$ in no-color-MLP frameworks, as modifications on zoomed-in Gaussian primitives are generally invisible in normal views. In the object training \textbf{UP-Mip} and \textbf{UP-3D-GS}, the upscaling loss is added directly.

%---------------------------------
\begin{figure*}[t!]
\centering
\includegraphics[width=1.\linewidth]{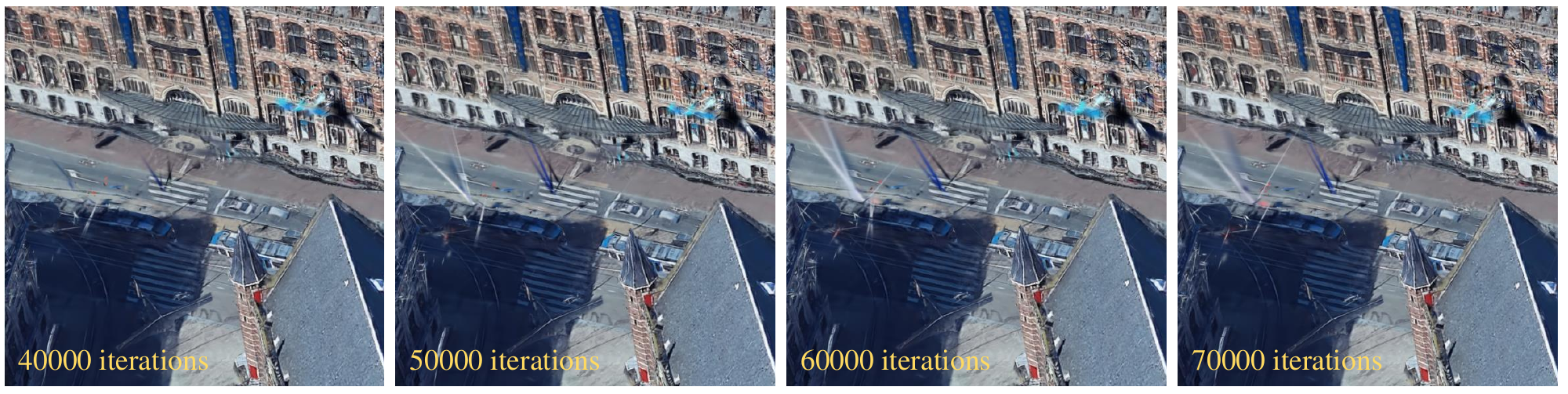}
% \vspace{-2em}
\caption{Artifacts of Octree-GS in zoomed-in views compared with different training iterations.}
\label{fig:time_arti_com}
\centering
\end{figure*}
%---------------------------------

\section{Experiments}
\subsection{Performance on Different Training Time} 
\label{sec:train_time_comp}
Based on our experiments and an analysis of the original 3D Gaussian Splatting (3D-GS) results~\cite{kerbl20233d}, we observe that beyond a certain performance threshold, further increasing the training time does not lead to improved performance. Specifically, the results from the original 3D-GS indicate that training for 70,000 iterations yields similar or even degraded performance compared to training for 30,000 iterations. In our study, we additionally trained the Octree-Scaffold-GS~\cite{ren2024octreegs} model for 80,000 iterations. Consistent with the observations from the original 3D-GS, the Octree-Scaffold-GS also encounters a performance bottleneck, as shown in Table~\ref{tab:time_comp_tnt}. Note that the time required to train for 55,000 iterations is equivalent to that of our \textbf{Octree-UP} while training for 70,000 iterations is comparable in time to our \textbf{Octree-UB}.

Moreover, extending the training time does not alleviate the zoomed-in artifacts. We trained Octree-Scaffold-GS on the multi-scale dataset BungeeNeRF for 70,000 iterations, which exceeds the training time required by our \textbf{Octree-UP-x4} method. Despite the prolonged training, we observed no improvement in performance metrics, and the zoomed-in artifacts persisted without mitigation; in fact, they were even exacerbated, as shown in Figure~\ref{fig:time_arti_com}. These results further demonstrate the effectiveness and necessity of our methods.

Through our experiments, we have found that this issue occurs because almost all regions exhibiting strong artifacts are not visible in the training dataset, and any modifications to the appearance of these Gaussian primitives are made during the early stages of training. After the scene reconstruction is relatively completed, these Gaussian primitives become occluded and are no longer visible from any training perspective, causing their features to remain unchanged during further training. However, the color MLP continues to evolve throughout the training process. As the color MLP in Octree-Scaffold-GS undergoes changes, the previous features are unable to generate the same colors with the modified MLP. Consequently, artifacts in these areas become exacerbated through further training. 

To further validate the effectiveness of our method, we conducted extended training of the Octree-3D-GS model on the BungeeNeRF dataset. With the increased training time, its artifacts did not worsen; however, they remained unchanged throughout the additional training. This observation is consistent with the findings reported in the original 3D-GS study~\cite{kerbl20233d}, where prolonged training similarly failed to enhance performance matrices and alleviate persistent artifacts.

\subsection{Results on Different Framework}
Since both \textbf{Octree-UP} and \textbf{Octree-UB} are based on the MLP-based Scaffold-GS framework, we additionally trained another version using the traditional Spherical Harmonics-based original 3D-GS to further validate the effectiveness of our methods on the Mip-NeRF360 dataset. The results are provided in Table~\ref{tab:up_or_3d_gs}. With the refined bootstrapping pipeline, the improvements in metrics are more pronounced than those achieved by the original Bootstrap-GS pipeline as demonstrated in \textbf{3D-GS-B}, where we use the same training configuration of \textbf{Bootstrap-GS} but different bootstrapping implementation. Consistent with our conclusion in Section~\ref{sec:main_imple}, training on Spherical Harmonics better leverages the potential of bootstrapping and upscaling, as each Gaussian primitive can be modified independently without the influence of the color prediction MLP. Note that the results of Bootstrap-GS are trained on our finetuned diffusion models with its original configuration.

\begin{table}[t!]
\centering
\renewcommand{\arraystretch}{1.05}
\setlength{\tabcolsep}{3pt}
\caption{Quantitative comparison of different frameworks on Mip-NeRF360~\cite{barron2022mipnerf360} datasets with our methods.}
\vspace{-3pt}
\label{tab:up_or_3d_gs}
%\resizebox{\linewidth}{!}
{
%\footnotesize
\small
\begin{tabular}{l|ccc}
\toprule
Method & PSNR($\uparrow$) & SSIM($\uparrow$) & LPIPS($\downarrow$) \\

\midrule
3D-GS~\cite{kerbl20233d} & 27.54 & 0.815 & 0.216\\

Octree-3D-GS \cite{ren2024octreegs} & 27.65 & 0.815 & 0.220 \\

Bootstrap-GS~\cite{gao2024bootstrap3dgs} 
& 28.32 & 0.823 & 0.209\\
\midrule

3D-GS-UP & 27.76 & 0.818 & 0.212  \\

Oct-3DGS-UP & 27.81 & 0.818 & 0.217 \\

3D-GS-B & 28.51 & 0.825 & \textbf{0.207}\\

3D-GS-UB & 28.59 & 0.826 & \textbf{0.207} \\

Oct-3DGS-UB & \textbf{28.71} & \textbf{0.828} & 0.211 \\

\bottomrule
\end{tabular}}
%\vspace{-2em}
\end{table}

\subsection{Ablation Study} 

%------------------------------------------
%\vspace{-2em}
\begin{table*}[htbp]
\centering
\renewcommand{\arraystretch}{1.05}
\setlength{\tabcolsep}{1pt}
\caption{Ablation studies on DeepBlending \cite{hedman2018deep} dataset. In the \#GS(k)/Mem matrix, entries marked with a ``-" indicate that the variations compared to the corresponding original versions are negligible, and \#GS(k)/Mem is also not the primary focus of these studies, these metrics do not provide meaningful comparisons.}
\vspace{-6pt}
%\resizebox{\linewidth}{!}
{
\small
%\footnotesize
\begin{tabular}{l|cccc}
\toprule
Dataset & \multicolumn{4}{c}{Deep Blending} \\
\begin{tabular}{c|c} Method & Metrics \end{tabular}  & ~PSNR $(\uparrow)$ ~ & SSIM $(\uparrow)$ ~ & LPIPS $(\downarrow)$ ~ & \#GS(k)/Mem  \\
\midrule
Octree-GS \cite{ren2024octreegs} &30.49 & 0.912 & 0.241 & 112/71.7M \\
Octree-UB & 31.23 & 0.919 & 0.237 & 117/75.6M  \\
\midrule
Octree-UB-raw & 30.62 & 0.913 & 0.240 & 116/74.2M  \\
Octree-UB-prompt-1 & 30.61 & 0.913 & 0.240 & -  \\
Octree-UB-prompt-2 & 30.52 & 0.911 & 0.241 & -  \\
Octree-UB-infer-1 & 31.23 & 0.919 & 0.237 & -  \\
Octree-UB-infer-n & 31.29 & 0.920 & 0.237 & -  \\
Octree-UB-var-1 & 31.16 & 0.918 & 0.238 & 115/73.7M  \\
Octree-UB-var-3 & 31.17 & 0.919 & 0.238 & 118/75.5M  \\
Octree-UB-uncropped ~& 31.11 & 0.916 & 0.239 & 120/77.1M \\
Octree-UB-start-15k & 31.38 & 0.921 & 0.236 & 122/79.4M  \\
Octree-UB-start-10k & 31.49 & 0.923 & 0.235 & 126/80.7M  \\
Octree-UB-loss+ & 31.31 & 0.920 & 0.237 & 122/79.4M  \\
Octree-UB-de-MLP & 31.32 & 0.919 & 0.237 & -  \\
\bottomrule
\end{tabular}
}
\label{tab:ablation_study_bungee}
\end{table*}
%------------------------------------------

%------------------------------------------
\begin{figure*}[t!]
\centering
\includegraphics[width=0.95\linewidth]{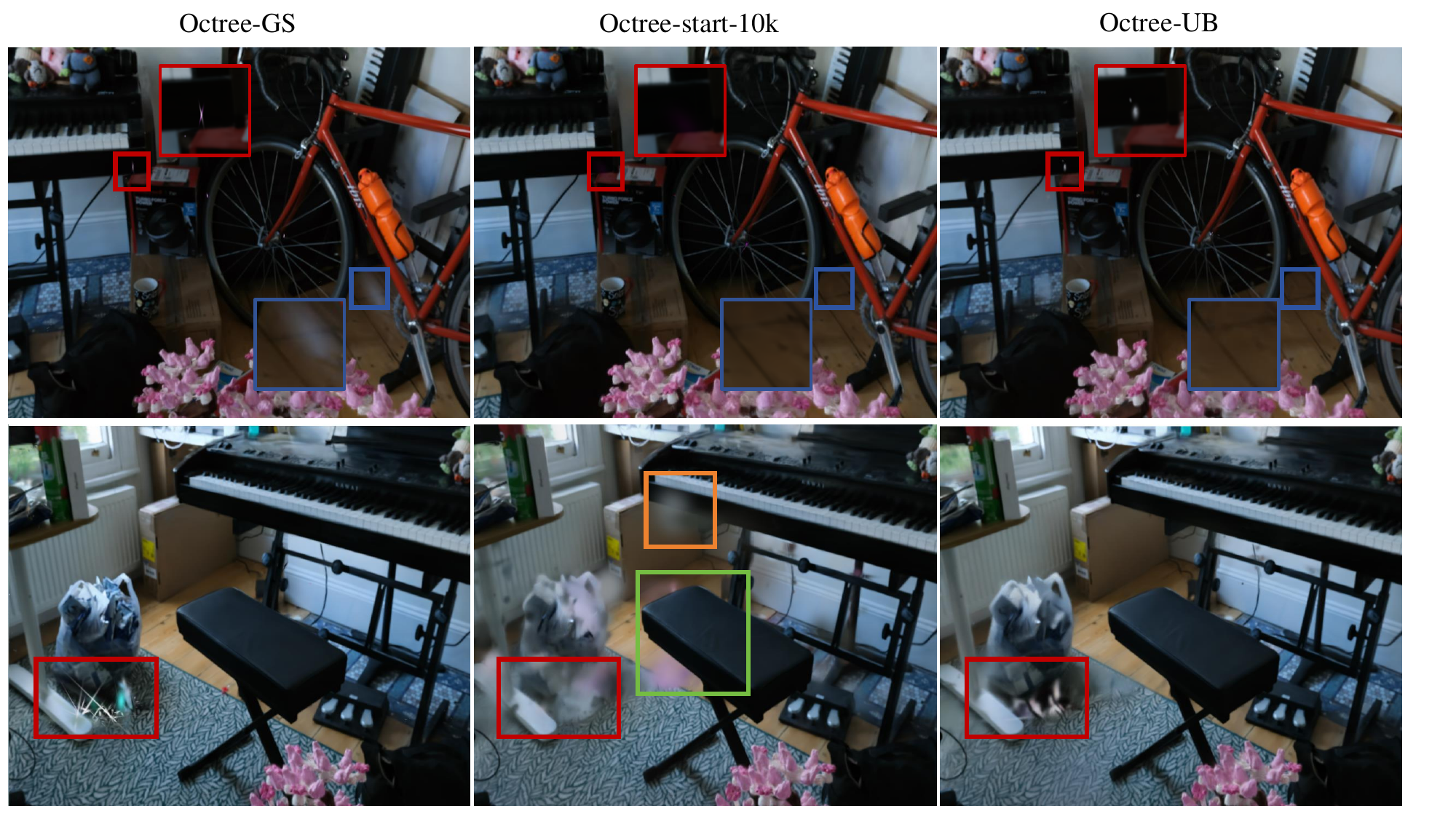}
% \vspace{-2em}
\caption{Artifacts of Mip-NeRF360 \textbf{Bonsai} in zoomed-in views compared with different frameworks.}
\label{fig:over_train_arti}
\centering
\end{figure*}
%------------------------------------------

Our ablation study primarily focuses on the configurations and implementations of the diffusion models within our framework. We present the results of these ablation studies on bootstrapping diffusion models in Table~\ref{tab:ablation_study_bungee}.

%\paragraph{Bootstrapping Ablation Study.}
\subsubsection{Bootstrapping Ablation Study}
In this context, \textbf{Octree-UB-raw} refers to the raw implementation of \textbf{SDXL-Turbo} without fine-tuning on each specific dataset. While this configuration shows slight overall improvements, the increments are inconsistent across different scenes. For instance, the results for the \textbf{Drjohnson} scene exhibit slight degradation, whereas the \textbf{Playroom} scene shows significant enhancement.

We also experimented with different textual prompts for conditional generation using the unfinetuned SDXL-Turbo, since after fine-tuning, we can discard the prompt guidance altogether. In \textbf{Octree-UB-prompt-1}, we used the prompt \textbf{``professional graph with fine-grained details"}, similar to the one employed in Bootstrap-GS~\cite{gao2024bootstrap3dgs}. This resulted in minimal changes in performance. However, in \textbf{Octree-UB-prompt-2}, where we added each scene's name in the first word of our original prompt, the results exhibited degradation. This underscores the necessity of fine-tuning the bootstrapping diffusion model, as the pretrained posterior of \textbf{SDXL-Turbo} may not align well with our datasets.

Further, we evaluated the impact of different inference time steps. \textbf{Octree-UB-infer-1} corresponds to our default Octree-UB configuration, performing inference in a single step. In contrast, \textbf{Octree-UB-infer-n} sets the number of inference steps to a fixed value of 100, thereby conducting multiple steps during inference. Although this approach yields a modest performance increase, it requires significantly more computational time. Therefore, a single inference step is sufficient for basic training.

We also tested the effect of varying the number of variants used in bootstrapping. \textbf{Octree-UB-var-1} utilizes one variant, while \textbf{Octree-UB-var-3} employs three variants. The results indicate that using only one variant may be insufficient to achieve the multi-view alignment required in Bootstrap-GS. Conversely, using three variants introduces excessive disturbance due to the random sampling of camera variants, which adversely affects the performance of the color MLP. Additionally, the uncropped bootstrapping training configuration \textbf{Octree-UB-uncropped} proved to be significantly more unstable than our cropped version.

Moreover, initiating bootstrapping at earlier stages of training can enhance performance in normal views, as shown in \textbf{Octree-UB-start-15k} and \textbf{Octree-UB-start-10k}, where bootstrapping begins at 15,000 and 10,000 iterations, respectively. However, this strategy is not universally beneficial for all scenes in zoomed-in views, particularly for certain scenes in the Mip-NeRF 360 dataset, as illustrated in Figure~\ref{fig:over_train_arti}. In Figure~\ref{fig:over_train_arti}(a), where the scenes are generally homogeneous, increased bootstrapping yields significantly better results in zoomed-in views compared to the original Octree-GS. However, in Figure~\ref{fig:over_train_arti}(b), where the scenes are largely collapsed in the original models, our techniques tend to produce overly smoothed results. In these cases, the diffusion models may lack sufficient information to accurately complete the images, since their finetuning is based on normal views where scenes are better reconstructed, as described in Bootstrap-GS~\cite{gao2024bootstrap3dgs}.

As a result, due to the averaging effect in the bootstrapping and upscaling loss functions, these scenes appear blurred and over-smoothed. While over-smoothing can mitigate sharp and chaotic noise—which is preferable to having prominent artifacts—the color MLP can exacerbate over-smoothing in regions that originally lacked significant artifacts. Specifically, in Figure~\ref{fig:over_train_arti}(b), \textbf{Octree-UB-start-10k} displays over-smoothed artifacts even in normal areas. This phenomenon aligns with our discussion in Section~\ref{sec:train_time_comp}, where Gaussian primitives modified at early stages become invisible in later stages, and the alterations introduced by the color MLP during bootstrapping lead to over-smoothed artifacts. In addition, we adjusted the loss weight \( \lambda_b = [0.25, 0.2] \) in \textbf{Octree-UB-loss+} based on \textbf{Octree-UB}, and this adjustment resulted in similar over-smoothed artifacts as those encountered in \textbf{Octree-UB-start-10k}.
%we observed that

Finally, in the configuration denoted as \textbf{Octree-UB-de-MLP}, we experimented with deactivating the MLP updating during bootstrapping iterations, building upon \textbf{Octree-UB-start-10k}. In this setup, only the features of Gaussian primitives~\cite{lu2023scaffold} are modified during bootstrapping, effectively eliminating the overly smoothed artifacts in normal regions caused by the color MLP. Although the training of the color MLP is affected and the normal-view matrices are slightly worse than those of \textbf{Octree-UB-start-15k}, we consider this a worthwhile trade-off to achieve more integrated zoomed-in views. However, in our main paper, we did not adopt this strategy because it increases the training time and we aimed to maintain fair comparisons with other frameworks without altering their training strategies. 

%\paragraph{Upsaling Ablation Study}
\subsubsection{Upsaling Ablation Study}
In this part, we have tried extensive configurations of upscale diffusion models and the loss values, especially on the multi-scale dataset BungeeNeRF. However, from the perspective of matrices, the differences are negligible, as shown in Table~\ref{tab:upscale_ablation}.

%------------------------------------------
%\vspace{-2em}
\begin{table*}[tbp]
\centering
\renewcommand{\arraystretch}{1.15}
\setlength{\tabcolsep}{1pt}
\caption{Ablation studies on the BungeeNeRF~\cite{xiangli2022bungeenerf} scenes.}
%\vspace{-6pt}
%\resizebox{\linewidth}{!}
{
%\footnotesize
\small
%\begin{tabularx}{0.8\textwidth}{l|ccc|ccc}
\begin{tabular}{l|cccc}
\toprule
Dataset & \multicolumn{4}{c}{BungeeNeRF} \\
\begin{tabular}{c|c}
\label{tab:upscale_ablation}
Method & Metrics \end{tabular}  & ~PSNR $(\uparrow)$ ~ & SSIM $(\uparrow)$ ~ & LPIPS$(\downarrow)$ ~ &  \#GS(k)/Mem \\
\midrule
Octree-GS \cite{ren2024octreegs} & 28.39 & 0.923 & 0.088& 1474/296.7M  \\

Octree-UPx2 & 28.43 & 0.923 & 0.088 & 1521/306.2M  \\

Octree-UPx4 & 28.48 & 0.924 & 0.087 & 1568/315.6M \\

\midrule
Octree-UPx2-l+ & 28.45 & 0.923 & 0.088 & 1523/306.6M  \\

Octree-UPx4-l+ & 28.43 & 0.923 & 0.088 & 1569/315.8M  \\

Octree-UPx2-var-1 & 28.41 & 0.923 & 0.087 & 1513/304.5M   \\
Octree-UPx4-var-1 & 28.46 & 0.924 & 0.088 & 1544/310.8M  \\

Octree-UPx2-var-3 & 28.49 & 0.925 & 0.087 & 1529/307.9M  \\

Octree-UPx4-var-3 & 28.52 & 0.925 & 0.086 & 1591/320.4M  \\

Octree-UPx2-uncropped ~ & 28.33 & 0.921 & 0.090 & 1545/311.1M  \\

Octree-UPx4-uncropped ~ & 28.27 & 0.921 & 0.091 & 1597/321.4M  \\

\bottomrule
\end{tabular}
}
\end{table*}
%------------------------------------------

In the configurations \textbf{Octree-UPx2-l+} and \textbf{Octree-UPx4-l+}, we decoupled the upscaling loss from the scaling factor \( \lambda_u \) and included it as an independent component in the loss function. We observed that the performance in normal views and the properties of Gaussian primitives remained largely unchanged, indicating that zoomed-in views exert minimal influence on normal-scale renderings. However, regarding Gaussian expansion, we found that the upscaling loss was excessively large even when \( \lambda_u \) was set to a very small value. Consequently, the duplication of Gaussian primitives had a magnitude similar to that of the scaled upscaling loss. As a result, the increase in the number of Gaussian primitives compared to \textbf{Octree-UPx2} and \textbf{Octree-UPx4} was minimal.

Nonetheless, augmenting the loss in Octree-Scaffold-GS led to more pronounced over-smoothed rendering results than those depicted in Figure~\ref{fig:over_train_arti}. We previously discussed the reasons for this in both Section~\ref{sec:train_time_comp} and the preceding paragraph. Even with numerous expanded views, most of the zoomed-in regions are not adequately modified during training. An excessively large upscaling loss disrupts the original distribution of the color MLP, thereby affecting the imperceptible Gaussian primitives that were initially not significantly divergent from the overall scene. Conversely, in the absence of a color MLP, we can directly incorporate the upscaling loss without applying a scaling factor.

After adjusting the number of variants used to generate random upscaling cameras, as described in Section~\ref{sec:implementaion_details}, we observed that increasing the number of constructed variants led to improved results. Since many Gaussian primitives in zoomed-in regions are only trained during the early stages (illustrated in Section~\ref{sec:train_time_comp}), introducing additional zoomed-in cameras helps to better integrate the entire scene from specific perspectives. Although the variations in evaluation metrics are not significant, the zoomed-in renderings can be substantially improved. The observed variations in the corresponding Gaussian primitives further validate our conclusion. Notice that, here, \( \lambda_u \) is very small and would not affect the overall performance of color MLP.

For uncropped upscaling configurations, namely \textbf{Octree-UPx2-uncropped} and \textbf{Octree-UPx4-uncropped}, the camera horizons are significantly expanded, resulting in a substantial increase in the expansion of Gaussian primitives. However, we do not recommend this approach. We found that when upscaling larger images, the diffusion models tend to generate more flexible details that are not permissible under our upscaling constraints. As a result, we observed noticeable performance degradation. Additionally, before performing large-scale upscaling, it is important to ensure that the code does not automatically embed watermarks in the generated images defaulted in \textbf{Stable Diffusion Configuration}, as this can also adversely affect the results.

\subsection{Per-scene Results} 
In Tables \ref{tab:results_tnt}-\ref{tab:lpips_bungee}, we list the metrics used in our evaluation in Sec.~\ref{sec:exp} across all considered methods and scenes.
% \vspace{-6pt}

%------------------------------------------
%\vspace{-2em}
\begin{table*}[htbp]
\centering
\renewcommand{\arraystretch}{1.05}
\setlength{\tabcolsep}{1pt}
\caption{Quantitative results for all scenes in the Tanks\&Temples \cite{knapitsch2017tanks} dataset.}
\vspace{-6pt}
%\resizebox{\linewidth}{!}
{
%\footnotesize
\small
\begin{tabular}{l|ccc|ccc}
\toprule
Dataset & \multicolumn{3}{c|}{Truck} & \multicolumn{3}{c}{Train} \\
\begin{tabular}{c|c}
\label{tab:results_tnt}
Method & Metrics \end{tabular}  & PSNR($\uparrow$) ~ & SSIM($\uparrow$) ~ & LPIPS$(\downarrow)$ ~ & PSNR($\uparrow$) ~ & SSIM($\uparrow$) ~ & LPIPS$(\downarrow)$ ~ \\
\midrule
3D-GS \cite{kerbl20233d} & 25.52 & 0.884 & 0.142  & 22.30 & 0.819 & 0.201 \\
Mip-Splatting \cite{yu2023mipsplat} ~ & 25.74 & 0.888 & 0.142  & 22.17 & 0.824 & 0.199  \\
Scaffold-GS \cite{lu2023scaffold} & 26.04 & 0.889 & 0.131  & 22.91 & 0.838 & 0.181  \\
Boot-GS \cite{gao2024bootstrap3dgs} & 26.23 & 0.891 & 0.139  & 23.47 & 0.835 & 0.187 \\
Octree-GS \cite{ren2024octreegs} & 26.24 & 0.894 & 0.122  & 23.11 & 0.838 & 0.184  \\
\midrule
Octree-UP & 26.21 & 0.894 & 0.122  & 23.34 & 0.841 & 0.182  \\
Octree-UB & \textbf{26.47} & \textbf{0.896} & \textbf{0.120}  & \textbf{23.83} & \textbf{0.846} & \textbf{0.179}  \\
\bottomrule
\end{tabular}
}
\end{table*}
%------------------------------------------

%------------------------------------------
%\vspace{-2em}
\begin{table*}[htbp]
\label{tab:results_db}
\centering
\renewcommand{\arraystretch}{1.05}
\setlength{\tabcolsep}{1pt}
\caption{Quantitative results for all scenes in the DeepBlending \cite{hedman2018deep} dataset.}
\vspace{-6pt}
%\resizebox{\linewidth}{!}
{
%\footnotesize
\small
\begin{tabular}{l|ccc|ccc}
\toprule
Dataset & \multicolumn{3}{c|}{Dr Johnson} & \multicolumn{3}{c}{Playroom} \\
\begin{tabular}{c|c} Method & Metrics \end{tabular}  & PSNR$(\uparrow)$ ~ & SSIM$(\uparrow)$ ~ & LPIPS$(\downarrow)$ ~ & PSNR$(\uparrow)$ ~ & SSIM$(\uparrow)$ ~ & LPIPS$(\downarrow)$ ~ \\
\midrule
3D-GS \cite{kerbl20233d} & 29.09 & 0.900 & 0.242 & 29.83 & 0.905 & 0.241 \\
Mip-Splatting \cite{yu2023mipsplat} ~& 29.08 & 0.900 & 0.241 & 30.03 & 0.902 & 0.245 \\
Scaffold-GS \cite{lu2023scaffold} & 29.73 & 0.910 & \textbf{0.235} & 30.83 & 0.907 & \textbf{0.242} \\
Boot-GS \cite{gao2024bootstrap3dgs} & \textbf{30.92} & \textbf{0.911} & 0.231 & \textbf{31.95} & 0.918 & \textbf{0.232} \\
Octree-GS \cite{ren2024octreegs} & 29.83 & 0.909 & 0.237 & 31.15 & 0.914 & 0.245 \\
\midrule
Octree-UP & 29.94 & 0.910 & 0.236 & 31.37 & 0.917 & 0.242 \\
Octree-UB & 30.64 & 0.915 & 0.234  & 31.82 & \textbf{0.921} & 0.240 \\
\bottomrule
\end{tabular}
}
\end{table*}
%------------------------------------------

\begin{table*}[htbp]
\renewcommand{\arraystretch}{1.1}
\setlength{\tabcolsep}{1pt}
\centering
\caption{PSNR ($\uparrow$) for all scenes in the Mip-NeRF360 \cite{barron2022mipnerf360} dataset.}
%\vspace{-6pt}
%\resizebox{\linewidth}{!}
{
%\footnotesize
\small
\begin{tabular}{l|ccccccccc}
\toprule
\begin{tabular}{c|c} Method & Scenes \end{tabular} & Bicycle & Bonsai &  Counter & Flowers & Garden & Kitchen & Room & Stump & Treehill \\
\midrule

3D-GS \cite{kerbl20233d} & 25.10 & 32.19 & 29.22 & 21.57 & 27.45 & 31.62 & 31.53 & 26.70 & 22.46 \\

Mip-Splatting \cite{yu2023mipsplat}~ & 25.13 & 32.56 & 29.30 & 21.64 & 27.43 & 31.48 & 31.73 & 26.65 & 22.60 \\

Scaffold-GS \cite{lu2023scaffold} & 25.19 & 33.22 & 29.99 & 21.40 & 27.48 & 31.77 & 32.30 & 26.67 & 23.08 \\

Boot-GS \cite{gao2024bootstrap3dgs} & \textbf{26.02} & 33.31 & 29.98 & 21.93 & \textbf{28.34} & 32.11 & 32.28 & 27.68 & 23.31  \\

Octree-GS \cite{ren2024octreegs} & 25.24 & 33.76 & 30.19 & 21.46 & 27.67 & 31.84 & 32.51 & 26.63 & 23.13 \\
\midrule

Octree-UP & 25.22 & 33.85 & 30.27 & 21.65 & 27.62 & 31.96 & 32.55 & 26.83 & 23.31 \\

Octree-UB & 25.61 & \textbf{33.98} & \textbf{30.63} & \textbf{22.14} & 27.93 & \textbf{32.19} & \textbf{32.85} & \textbf{27.72} & \textbf{23.77} \\
\bottomrule

\end{tabular}
}
\end{table*}

%\vspace{-0.5em}
\begin{table*}[htbp]
\renewcommand{\arraystretch}{1.1}
\setlength{\tabcolsep}{1pt}
\centering
\caption{SSIM ($\uparrow$) for all scenes in the Mip-NeRF360 \cite{barron2022mipnerf360} dataset.}
%\vspace{-6pt}
%\resizebox{\linewidth}{!}
{
%\footnotesize
\small
\begin{tabular}{l|ccccccccc}
\toprule
\begin{tabular}{c|c} Method & Scenes \end{tabular} & Bicycle & Bonsai &  Counter & Flowers & Garden & Kitchen & Room & Stump & Treehill \\
\midrule
3D-GS \cite{kerbl20233d} & 0.747 & 0.947 & 0.917 & 0.600 & 0.861 & 0.932 & 0.926 & \textbf{0.773} & 0.636 \\
Mip-Splatting \cite{yu2023mipsplat}~ & 0.747 & 0.948 & 0.917 & 0.601 & 0.861 & 0.933 & 0.928 & 0.772 & 0.639 \\
Scaffold-GS \cite{lu2023scaffold} & 0.751 & 0.952 & 0.922 & 0.587 & 0.853 & 0.931 & 0.932 & 0.767 & 0.644 \\
Boot-GS \cite{gao2024bootstrap3dgs} & \textbf{0.763} & 0.951 & 0.922 & \textbf{0.608} & \textbf{0.868} & 0.937 & 0.935 & \textbf{0.784} & 0.645 \\
Octree-GS \cite{ren2024octreegs} & 0.755 & 0.955 & 0.925 & 0.595 & 0.861 & 0.933 & 0.936 & 0.766 & 0.641 \\
\midrule
Octree-UP & 0.755 & 0.956 & 0.924 & 0.597 & 0.861 & 0.932 & 0.935 & 0.769 & 0.645 \\
Octree-UB & 0.759 & \textbf{0.957} & \textbf{0.930} & 0.604 & 0.863 & \textbf{0.938} & \textbf{0.938} & \textbf{0.784} & \textbf{0.650} \\
\bottomrule

\end{tabular}
}
\end{table*}

%\vspace{-0.5em}
\begin{table*}[htbp]
\renewcommand{\arraystretch}{1.1}
\setlength{\tabcolsep}{1pt}
\centering
\caption{LPIPS ($\downarrow$) for all scenes in the Mip-NeRF360 \cite{barron2022mipnerf360} dataset.}
\vspace{-6pt}
%\resizebox{\linewidth}{!}
{
%\footnotesize
\small
\begin{tabular}{l|ccccccccc}
\toprule
\begin{tabular}{c|c} Method & Scenes \end{tabular} & Bicycle & Bonsai &  Counter & Flowers & Garden & Kitchen & Room & Stump & Treehill \\
\midrule

3D-GS \cite{kerbl20233d} & 0.243 & 0.178 & 0.179 & 0.345 & 0.114 & 0.117 & 0.196 & 0.231 & 0.335 \\
Mip-Splatting \cite{yu2023mipsplat}~ & 0.245 & 0.178 & 0.179 & 0.347 & 0.115 & 0.115 & 0.192 & 0.232 & \textbf{0.334} \\
Scaffold-GS \cite{lu2023scaffold} & 0.247 & 0.173 & 0.177 & 0.359 & 0.130 & 0.118 & 0.183 & 0.252 & 0.338 \\
Boot-GS \cite{gao2024bootstrap3dgs} & 0.236 & 0.169 & 0.174 & 0.341 & 0.110 & 0.114 & 0.191 & \textbf{0.225} & 0.327 \\
Octree-GS & \textbf{0.235} & 0.164 & 0.169 & 0.347 & 0.116 & 0.115 & 0.172 & 0.250 & 0.360 \\
\midrule

Octree-UP & 0.236 & 0.164 & 0.168 & 0.345 & 0.116 & 0.117 & 0.173 & 0.247 & 0.356 \\

Octree-UB & \textbf{0.235} & \textbf{0.161} & \textbf{0.165} & \textbf{0.341} & \textbf{0.114} & \textbf{0.113} & \textbf{0.171} & 0.242 & 0.354 \\
\bottomrule

\end{tabular}
}
\end{table*}

%------------------------------------------
%\vspace{-2em}
\begin{table*}[htbp]
\renewcommand{\arraystretch}{1.1}
\setlength{\tabcolsep}{1pt}
\centering
\caption{PSNR ($\uparrow$) for all scenes in the BungeeNeRF \cite{xiangli2022bungeenerf} dataset.}
%\vspace{-6pt}
%\resizebox{\linewidth}{!}
{
%\footnotesize
\small
\begin{tabular}{l|ccccccccc}
\toprule
\begin{tabular}{c|c} Method & Scenes \end{tabular}  & Amsterdam & Barcelona & Bilbao & Chicago & Hollywood & Pompidou & Quebec & Rome\\
\midrule
3D-GS \cite{kerbl20233d} & 27.75 & 27.55 & 28.91 & 28.27 & 26.25 & 27.16 & 28.86 & 27.56 \\

Mip-Splatting \cite{li2024mipmapgs}~ & \textbf{28.16} & 27.72 & 29.13 & 28.28 & 26.59 & 27.71 & 29.23 & 28.33 \\

Scaffold-GS \cite{lu2023scaffold} & 27.82 & 28.09 & 29.20 & 28.55 & 26.36 & \textbf{27.72} & 29.29 & 28.24 \\

Octree-GS \cite{ren2024octreegs} & 28.16 & 28.40 & 29.39 & 28.86 & \textbf{26.76} & 27.46 & 29.46 & 28.59 \\
\midrule
Octree-UPx2 & 28.14 & 28.32 & \textbf{29.51} & 28.94 & 26.71 & \textbf{27.64} & 29.51 & 28.55 \\

Octree-UPx4 & 28.18 & \textbf{28.43} & 29.47 & \textbf{29.11} & 26.71 & 27.49 & \textbf{29.62} & \textbf{28.68} \\
\bottomrule
\end{tabular}
}
\end{table*}
%------------------------------------------

%------------------------------------------
%\vspace{-2em}
\begin{table*}[htbp]
\renewcommand{\arraystretch}{1.1}
\setlength{\tabcolsep}{1pt}
\centering
\caption{SSIM ($\uparrow$) for all scenes in the BungeeNeRF \cite{xiangli2022bungeenerf} dataset.}
%\vspace{-6pt}
%\resizebox{\linewidth}{!}
{
%\footnotesize
\small
\begin{tabular}{l|ccccccccc}
\toprule
\begin{tabular}{c|c} Method & Scenes \end{tabular} & Amsterdam & Barcelona & Bilbao & Chicago & Hollywood & Pompidou & Quebec & Rome\\
\midrule

3D-GS \cite{kerbl20233d} & 0.918 & 0.919 & 0.918 & 0.932 & 0.873 & 0.919 & 0.937 & 0.918 \\

Mip-Splatting \cite{yu2023mipsplat}~ & 0.918 & 0.919 & 0.918 & 0.930 & 0.876 & 0.923 & 0.938 & 0.922 \\

Scaffold-GS \cite{lu2023scaffold} & 0.914 & 0.923 & 0.918 & 0.929 & 0.866 & 0.926 & 0.939 & 0.924 \\

Octree-GS \cite{ren2024octreegs} & \textbf{0.922} & 0.928 & 0.923 & 0.935 & \textbf{0.884} & 0.925 & 0.942 & 0.930 \\
\midrule

Octree-UPx2 & \textbf{0.922} & 0.927 & \textbf{0.924} & 0.934 & \textbf{0.884} & \textbf{0.927} & 0.942 & 0.930 \\

Octree-UPx4 & \textbf{0.922} & \textbf{0.929} & 0.923 & \textbf{0.936} & \textbf{0.884} & 0.925 & \textbf{0.944} & \textbf{0.931} \\
\bottomrule
\end{tabular}
}
\end{table*}
%------------------------------------------

%------------------------------------------
%\vspace{-2em}
\begin{table*}[htbp]
\renewcommand{\arraystretch}{1.1}
\setlength{\tabcolsep}{1pt}
\centering
\caption{LPIPS ($\downarrow$) for all scenes in the BungeeNeRF \cite{xiangli2022bungeenerf} dataset.}
%\vspace{-6pt}
%\resizebox{\linewidth}{!}
{
%\footnotesize
\small
\begin{tabular}{l|ccccccccc}
\toprule
\begin{tabular}{c|c}
\label{tab:lpips_bungee}
Method & Scenes \end{tabular} & Amsterdam & Barcelona & Bilbao & Chicago & Hollywood & Pompidou & Quebec & Rome\\
\midrule

3D-GS \cite{kerbl20233d} & 0.092 & 0.082 & 0.092 & 0.080 & 0.128 & 0.090 & 0.087 & 0.096 \\

Mip-Splatting \cite{yu2023mipsplat}~ & 0.094 & 0.082 & 0.095 & 0.081 & 0.130 & 0.087 & 0.087 & 0.093 \\

Scaffold-GS \cite{lu2023scaffold} & 0.102 & 0.078 & \textbf{0.090} & 0.08 & 0.157 & \textbf{0.082} & \textbf{0.080} & 0.087 \\

Our-Scaffold-GS & \textbf{0.090} & 0.071 & 0.091 & 0.077 & \textbf{0.128} & 0.089 & 0.081 & 0.080 \\

\midrule

Octree-UPx2 & 0.091 & 0.072 & \textbf{0.090} & 0.077 & \textbf{0.128} & 0.088 & 0.081 & 0.080 \\

Octree-UPx4 & 0.091 & \textbf{0.070} & 0.091 & \textbf{0.075} & \textbf{0.128} & 0.089 & \textbf{0.080} & \textbf{0.079} \\
\bottomrule
\end{tabular}
}
\end{table*}
%------------------------------------------


\end{document}